# Function, Complexity and Thermodynamics in Adaptive and Intelligent Soft-Matter Systems: An Information-Theoretical Conceptual Framework

George S. Attard

*School of Chemistry and Chemical Engineering, University of Southampton, Southampton SO17 1BJ, UK and Division of Physical Chemistry, Box 124, SE-221 00 Lund, Sweden.*

June 2026

(e-mail: gza@soton.ac.uk)

**Abstract.**

The terms *responsive*, *adaptive* and *intelligent* are widely used in the context of materials science and soft matter, but are qualitative in nature. There is currently no quantitative basis on which systems from different areas of materials science can be compared with one another on a common axis. This paper proposes a conceptual framework intended to address this gap. The approach taken is to treat any stimulus-coupled material as an information channel. The three classes are distinguished by the conditioning structure of the channel kernel: a memoryless map $p(y|x)$ (responsive), a state-conditioned map $p(y|x, s)$ (adaptive), and a feedback-modified channel $p(y_t|x_t, X_{t-1}, Y_{t-1})$ (intelligent). Three information-theoretic complexity metrics are defined on this basis: configurational diversity $I_1$, functional selectivity $I_2$, and stimulus–response information transfer $I_3$. Treating the material as the channel introduces a constraint absent from classical channel theory: the kernel and the physical substrate are the same object. This leads to a heuristic non-monotonic relationship between internal complexity and realised information transfer, together with an associated optimal internal complexity $N^*$ set by transmission efficiency, stimulus energy and thermal noise. A two-plane benchmarking scheme is proposed: a dynamic plane pairing the volumetric information rate $\frac{\dot{I}_3}{V}$ with power density $P$ against the Landauer–Bérut floor, and a companion static plane $(I_1, I_2)$. Sixteen representative systems spanning synthetic soft matter, biology and hard-matter architectures are placed on these coordinates. Within the stated uncertainties they appear to separate into broad bands above the Landauer–Bérut benchmark: approximately $10^{18}$–$10^{20}$× for soft matter and shape-memory alloys, $10^{10}$–$10^{16}$× for silicon digital and electromechanical devices, $10^{9}$–$10^{10}$× for memristor neuromorphic hardware, and $10^{5}$–$10^{8}$× for evolved biology, with an endoplasmic-reticulum lipid network as an intermediate case near $10^{11}$×. Each placement carries an estimated uncertainty of at least one to two decades on each axis, arising from the estimation assumptions and from the heuristic basis of parts of the framework; the band ordering appears robust to this uncertainty. The gap between synthetic soft matter and biology is tentatively attributed to the per-element substrate energy scale (of order 1–10 $k_BT$ versus $10^4$–$10^5$ $k_BT$). The proposed framework establishes the foundations that may lead to validated design rules. As an illustration of this, three routes are proposed through which the separation of soft matter systems from the Landauer benchmark might be reduced.

*Keywords* — adaptive soft matter, intelligent soft matter, complexity metrics, design space, information theory.

## 1. Introduction

Over the past two decades the language used to classify stimulus-coupled soft matter has evolved in step with the behaviours being reported. The earliest and still largest class, stimuli-responsive materials, comprises systems in which an environmental cue (e.g. temperature, pH, light, an electric field) drives a reversible change in conformation, assembly or a bulk property, with the material returning to its initial state once the trigger is removed [1]. Walther subsequently articulated a progression beyond this class, distinguishing adaptive materials, whose input–output relationship is history-dependent, from interactive materials, in which self-organising components communicate and respond collectively to a changing environment [2]. The most ambitious end of the spectrum has been framed as intelligent matter, in which 'intelligence' results from the integration of sensing, actuation and information processing within the substrate itself [3]. Baulin and colleagues set out a vision of embodied intelligence defined by four capabilities that a material must possess: perception/sensing, memory and learning, actuation, and decision-making/communication [4]. The field has thus converged, loosely, on a responsive → adaptive → intelligent continuum.

Soft matter systems are becoming increasingly sophisticated. Adamska *et al*. describe a single naphthalene-diimide component that reconfigures into five topologically distinct chiral assemblies under temperature, solvent, concentration and guest-molecule triggers, presenting this as evidence of the adaptability"of non-covalent assemblies whose functions adjust to external signals [5]. Ranganath and Maity review feedback-controlled reaction networks that confer homeostasis on soft materials: self-regulating, self-correcting and self-monitoring systems that preserve an internal state against cyclic external agitation [6]. Chiolerio *et al.* report learning in a colloidal suspension of polyaniline nanorods, which under AC stimulation develops programmable conductive pathways and synapse-like plasticity, allowing it to memorise configurational information [7]. Each set of authors characterises their system in the vocabulary of adaptiveness, memory or learning native to their subfield.

This terminological richness can be problematic. The definitions are qualitative and, as Walther [2] and Baulin et al. [4] both note, inconsistently applied. The central issue is that there is no agreed and measurable basis on which different material systems (e.g. a multistate supramolecular assembly or a learning colloid) can be placed on a common axis, still less compared with biological systems or subsystems. Existing complexity measures do not fill this gap, because they were built for other purposes (e.g. molecular-structural indices quantify the intricacy of a single molecular graph) [15], and statistical-complexity measures characterise the computational structure of a time series [21], but neither couples a material's function to its internal architecture or to the thermodynamic cost of operation, and neither is constructed to span material classes. The absence of such a quantitative framework is the principal obstacle to a design methodology: without metrics that are comparable across systems, one cannot identify where a given material sits on the continuum, what limits its performance, or which structural changes would move it toward more adaptive behaviour. Design rules require a measurable design space.

The approach developed in this paper is to treat any stimulus-coupled material as an information channel: when a material maps a stimulus $X$ onto a response $Y$, its functional behaviour is the input–output map, and the information it transmits is quantified by the mutual information $I(X;Y)$ [9]. This information-theoretic perspective is class-agnostic: it applies equally to a responsive gel, an adaptive network and a biological system. It thus provides a common, quantitative axis that the qualitative terminology lacks. It is also the natural route to a thermodynamic grounding, since information transmission and erasure carry a physical cost set by Landauer's principle [25]. This allows functional capacity to be linked to power density. On this basis the responsive–adaptive–intelligent continuum can be recast as a hierarchy of channels of increasing internal complexity, and a tentative quantitative design space for adaptive and intelligent soft matter can be constructed. The framework proposed here addresses three coupled questions. (i) For a given internal architecture, what sets the upper bound on $I(X;Y)$, and hence a material's functional richness? (ii) What determines the fraction of that bound realised under operating conditions? (iii) How do both scale with the internal complexity of the architecture?

## 2. The responsive–adaptive–intelligent continuum as a hierarchy of channels

The three classes are distinguished by the conditioning structure of the input–output channel kernel $p(y|x,\cdot)$ (Fig. 1). The term *kernel* denotes the conditional probability law that maps the input random variable to the output random variable; it is the channel itself, not a sample of its behaviour.

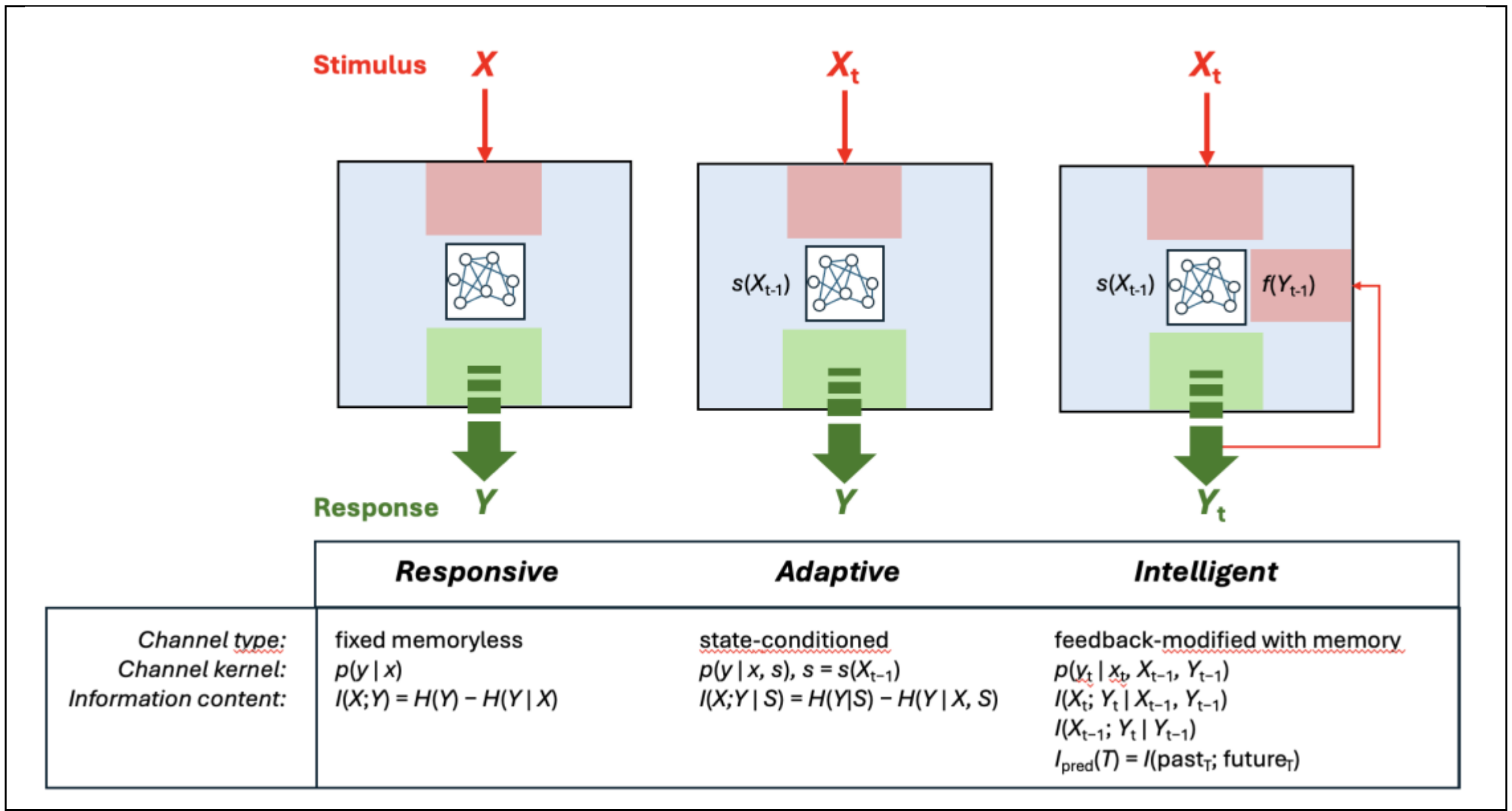

**Figure 1**. Schematic representation of the conceptual framework developed in this paper describing the properties of responsive, adaptive and intelligent soft matter continuum as a hierarchy of channel kernels of increasing conditioning complexity..

### *Responsive matter: a fixed memoryless channel*

A responsive material implements a time-invariant, history-independent map from $X$ to $Y$. The channel kernel is $p(y|x)$ and the functional information content (bits) is

$$I(X;Y) = H(Y) - H(Y|X) . \quad (1)$$

A poly($N$-isopropylacrylamide) (PNIPAM) hydrogel below its lower critical solution temperature (LCST) is the canonical example: temperature $X$ maps to swelling state $Y$ through a kernel determined by polymer chemistry and crosslink density and unchanged by prior history [5]. The channel is memoryless on operational timescales. For a responsive channel the upper bound on $I(X;Y)$ is the channel capacity $C = max_{p(x)} I(X;Y)$ [9].

### *Adaptive matter: a state-conditioned channel*

An adaptive material's kernel depends on an internal state $S$ that itself depends on input history. The channel is $p(y|x,s)$, $s = s(X_{t-1})$, with functional metric

$$I(X;Y|S) = H(Y|S) - H(Y|X,S) . \quad (2)$$

$I(X;Y)$ is not the natural measure here because it averages over the internal-state distribution and conflates the channel with a random variable. A glucose-responsive insulin-release matrix is adaptive in this sense: the release kernel depends on current swelling state and bound-glucose loading, set by recent glucose history. The channel is no longer time-invariant but remains passive: the kernel adjusts in response to inputs but does not modify itself in response to outputs.

### *Intelligent matter: a feedback-modified channel with memory*

The use of *intelligent* here is an operational engineering definition, not a cognitive one. An intelligent material's kernel depends on the history of inputs *and* on the history of its own outputs, as $Y$ is fed back to the channel:

$$p(y_t|x_t, X_{t-1}, Y_{t-1}) . \quad (3)$$

Three information-theoretic objects characterise such a channel.

(i) *Conditional mutual information* $I(X_t; Y_t|X_{t-1}, Y_{t-1})$, the per-step novel information transferred from input to output given the history.

(ii) *Transfer entropy* [10],

$$T_{X\to Y} = I(X_{t-1}; Y_t|Y_{t-1}) = H(Y_t|Y_{t-1}) - H(Y_t|Y_{t-1}, X_{t-1}) . \quad (4)$$

$T_{X\to Y}$ is the directed information rate from $X$ to $Y$, isolating the causal contribution of $X$ beyond what $Y$ predicts from its own past. The data-processing inequality [8] gives $T_{X\to Y} \leq I(X_{t-1}; Y_t)$, with inequality when $Y$ has internal memory.

(iii) *Predictive information* [12],

$$I_{pred}(T) = I(past_T; future_T) , \quad (5)$$

the mutual information between past and future segments of the output stream. $I_{pred}(T)$ grows as log $T$ for systems with a finite-parameter internal model and as a fractional power of $T$ for non-parametric ones; this growth law is a model-free signature of the complexity of the dynamics implementing the feedback.

The thermodynamic significance of eqn. (3) is that feedback turns mutual information into a thermodynamic resource. Sagawa and Ueda [13] showed that the second law of thermodynamics generalises, under feedback control, to

$$\langle W \rangle \geq \Delta F - k_B T\,(ln\, 2)\,\langle I \rangle\,, \qquad (6)$$

Where $\langle W \rangle$ is the ensemble-veraged work, $\langle I \rangle$ is the average mutual information (in bits) acquired by the feedback controller. The feedback-modified channel of an intelligent material is therefore not merely statistically different from the adaptive case; it has a different thermodynamic budget.

A number of soft-matter systems appear to satisfy eqn. (3), but it is not yet clear whether they are intelligent or adaptive [S8]. Living systems (bacterial chemotaxis, neural circuits, regulated metabolism) unambiguously meet the criteria for intelligent soft matter and have been characterised in the appropriate information-theoretic terms [14].

## 3. Existing complexity metrics

Complexity metrics are useful because they would provide a standard basis for placing a system on the responsive-to-intelligent continuum, for comparing soft matter against hard-matter electronics or biology, and for quantifying operational efficiency. A metric usable across the continuum should satisfy four requirements. (R1) *Dimensional consistency*: a single set of units in which all classes can be expressed. (R2) *Common reference*: definitions of “no complexity” and “maximum complexity” that survive the move between classes. (R3) *Thermodynamic coupling*: a route to absolute units (joules, watts) so the metric can be benchmarked against fundamental limits. (R4) *Scale-bridging*: definability at molecular, network and macroscopic scales without rederivation. The dominant existing metrics are assessed against these requirements below.

***Molecular-scale structural indices***. The Bertz index [15] and the Böttcher additive index [16] are graph-theoretic measures of bond-topology and stereochemistry information in a single molecule. The molecular assembly index MA [17] counts the minimum number of bond-forming steps required to build a molecule from atoms and is uniquely measurable through MS/MS fragmentation. These satisfy R1 (for entropy-based variants) and partially R2, but fail R3 and R4: there is no general thermodynamic coupling for a Bertz or Böttcher score, and one cannot infer a responsive system's dissipation rate from the structural complexity of its molecules.

***Network-scale invariants***. Chemical reaction network theory provides the stoichiometric rank, deficiency $\delta$ and cyclomatic number $\gamma$ as algebraic invariants of a reaction graph [18]. The deficiency is $\delta = n - \ell - s$, where *n* is the number of distinct complexes (the reactant and product groupings that appear in the network), $\ell$ is the number of linkage classes (connected components of the reaction graph) and $s$ is the rank of the stoichiometric subspace; δ is a non-negative integer measuring how far the network departs from the structural condition under which mass-action kinetics admits a unique, stable equilibrium. These give exact necessary conditions for multistationarity and oscillation (the Deficiency Zero Theorem) but are integer-valued and not expressible in bits, failing R1. They also lack inherent thermodynamic coupling: a network with δ = 1 can be close to equilibrium (low dissipation) or far from it (high dissipation), and the invariants do not distinguish the cases.

***Dynamical-scale measures***. The Lyapunov spectrum, the Kolmogorov–Sinai (KS) entropy $h_{KS} = \Sigma\lambda_i^+$ (bits·s$^{-1}$) and the Kaplan–Yorke dimension are the established characterisations of attractor dynamics [19]. The $\lambda_i$ are the Lyapunov characteristic exponents, which define the asymptotic mean exponential rates at which initially nearby trajectories on the attractor separate (positive $\lambda_i$) or converge (negative) along the *i*-th principal direction. the notation $\Sigma\lambda_i^+$ denotes the sum over the positive exponents only, which by the Pesin identity equals the KS entropy [16]. The KS entropy is the closest existing metric to an information-processing rate and is dimensionally compatible with the Landauer coupling (R3). However, it measures the rate of information *generation* by the intrinsic dynamics, not the rate of information *transfer* from a stimulus to a response, so it is mismatched to the channel-centred formulation. A material with high $h_{KS}$ can be a noisy non-responder; one with low $h_{KS}$ can be a highly informative deterministic switch.

***Compositional complexity measures***. The effective complexity of Gell-Mann and Lloyd [20] and the statistical complexity of Crutchfield and Young [21] both extract the regularity component of an entropy budget, i.e. the part attributable to structure rather than randomness. Both satisfy R1 (bits) and R2 (vanishing at both entropy extremes). However, neither has a general thermodynamic coupling (R3 partial), and neither is operationally easy to compute for stimulus–response channels (R4 partial). They are useful diagnostically but not as design metrics.

***The count-axis (D) metric*** A simple metric useful for order-of-magnitude differentiation is $D = log_2(count\ of\ distinguishable\ functional\ elements)$, with counts ranging from operational states of a hydrogel ($D \approx 7 - 10$) to the cells of a human ($D \approx 47$). The count is a useful zeroth-order positioning axis for functional complexity, resting on the heuristic that a system with larger $D$ offers more options for "function" (here understood as the external selection of one or more material properties or behaviours in an application context). While $D$ captures the order-of-magnitude separation between soft-matter, organelle and organismal complexity, its definition in terms of functional elements is open to multiple interpretations, and so is a qualitative metric at best.

## 4. The ($I_1$, $I_2$, $I_3$) complexity formulation

Three information-theoretically distinct primitives are proposed that meet requirements R1 to R4., Each metric is in bits, each with explicit boundary behaviour and each attached to a different design question. The derivations follow Shannon [9], Cover and Thomas [11] and the functional-information construction of Hazen, Griffin, Carothers and Szostak [22].

**Setup.** Let $\mathcal{S}$ be the set of microstates accessible under specified macroscopic constraints. Three choices fix the downstream complexity measures: the cardinality $|\mathcal{S}| = N$ (or the coarse-grained phase-space volume for continuous systems), the empirical occupancy $p$ on $\mathcal{S}$, and a reference distribution $u$, usually uniform with $u_i = 1/N$. The maximum-entropy reference is $H_{max} = log_2 N$, with $H(p) \leq H_{max}$ and equality iff $p = u$. The identity [23]

$$D_{KL}(p \parallel u) = H_{max} - H(p) \quad (7)$$

shows that the entropy deficit is itself a Kullback–Leibler (KL) divergence. The relevance of this identity is threefold. First, it gives the deficit a sign and a zero: a KL divergence is non-negative and vanishes only when $p = u$, so the gap between the realised entropy and its maximum is a proper measure of how far the occupancy departs from the disordered reference, not merely an arbitrary difference of two entropies. Second, it makes $I_n$ information-geometrically meaningful: the deficit is a directed "distance" in the space of distributions, with the convexity and additivity properties of a divergence, so configurational order can be compared across systems on a common, dimensionless footing. Third, it underwrites cross-class comparison: because KL divergences are invariant under refinement of the state-space partition, the deficit computed at one coarse-graining is consistent with that computed at another, which is the property the framework relies on when placing soft matter, organelles and devices on the same axis. For continuous state spaces $H_{max}$ is replaced by $log_2(Vol/\Delta^d)$ at coarse-graining resolution $\Delta$; differences in $H$ and all KL divergences are resolution-invariant, the property that permits cross-class comparison.

### $I_1$: configurational diversity

$$I_1 \equiv H(p) = -\Sigma p_i log_2 p_i \quad [bits]. \quad (8)$$

$I_1$ is the Shannon entropy of the empirical microstate occupancy (the bits of uncertainty about which microstate the material occupies). The boundary behaviour is: $I_1 \to 0$ for a material locked in a single state; $I_1 \to H_{max}$ for fully randomised occupancy and $I_1 = 1$ bit for a clean two-state switch with equal populations. $I_1$ is the irreducible Shannon primitive. Every other Shannon-derived quantity decomposes into entropies of marginal and joint distributions.

The discriminative power of $I_1$ for cross-class comparison depends on the coarse-graining at which microstates are defined. At the level of operational macrostates (swollen/collapsed, bound/unbound) all classes score $I_1$ in the range 1–4 bits, while at the level of per-chain or per-protein conformational entropy the same systems separate by several decades. The choice of coarse-graining is therefore part of the metric specification and must be stated explicitly (see SI S4 for the standardised procedure adopted here).

### $I_2$: functional selectivity

Throughout this paper, *function* is used to mean a property, or combination of properties, of a material that is selected by an external user for use in an application context, for example a sharp aqueous LCST at a useful temperature, a high binding selectivity for a target analyte, or a large reversible actuation strain. Function in this sense is not an intrinsic attribute of the material in isolation; it is defined jointly by the material and the use to which it is put. A configuration is "functional" when the selected property reaches a stated threshold, and $I_2$ quantifies (in bits) how rare such configurations are within a defined design space.

$$I_2 \equiv -log_2 F(E^*)\ , \qquad (9)$$

where $F(E^*) = M(E^*)/N$ is the random-search probability that a uniformly drawn configuration satisfies the functional threshold $\varphi(s) \geq E^*$ [22]. Here $M(E^*)$ is the number of distinct material configurations (compositions, sequences, morphologies or device parameter sets) whose degree of function $\varphi(s)$ (the measured value of the selected property) reaches or exceeds the threshold $E^*$, and $N$ is the size of the design space searched. $M(E^*)$ is therefore a direct count of how many ways the chosen material chemistry can be arranged so as to deliver at least the target level of the application-relevant property. Raising the threshold shrinks $M$ and increases $I_2$. $I_2$ is the number of bits needed to specify *any* configuration meeting the threshold from the design space. It measures the rarity of functional configurations and is independent of the empirical occupancy $p$. Its boundary behaviour is: $I_2 = 0$ when the function is generic ($F = 1$) and $I_2 \rightarrow log_2 N$ when the functional configuration is essentially unique. $I_2$ is the natural axis for inverse design but its heuristic basis currently limits its quantitative reach.

Hazen *et al.* reported $I_2 \approx$ 30–80 bits for biological RNA aptamers [22]. Comparable values for synthetic stimulus-responsive polymers have not been systematically tabulated; order-of-magnitude estimates follow from setting $F(E^*)$ equal to the fraction of a stated reference chemistry space whose members exhibit the target response above a useful threshold. The absence of community-agreed conventions for the reference space and threshold limits means that $I_2$ provides only a qualitative placement of systems. SI S5 sets out the two construction methods (combinatorial counting; design-specification bits) and the convention adopted for each system.

### $I_3$: input–output information transfer

For a responsive system $I_3 \equiv I(X;Y)$; for an adaptive system $I_3 \equiv I(X;Y|S)$; for an intelligent system $I_3 \equiv T_{X\rightarrow Y}$, the transfer entropy (eqn. 4). In all three cases $I_3$ has units of bits per channel use, with a rate version $\dot{I}_3 = I_3 \cdot \nu$ in bits·s$^{-1}$, where ν is the operational cycling frequency. Its boundary behaviour is: $I_3 = 0$ for a non-responsive material and $I_3 \rightarrow min\{H(X), H(Y)\}$ for a deterministic, perfectly distinguishable response. $I_3$ is bounded above by the channel capacity of the underlying physical channel and below by the data-processing inequality once the signal has passed through internal elements.

This paper foregrounds $I_3$ as the primitive that maps most directly onto experimental data and relies least on community-agreed conventions. The rate $\dot{I}_3$ is the only one of the three that admits a volumetric form (bits·s$^{-1}$·m$^{-3}$) dimensionally matched to a volumetric power density. This matching is what makes $\dot{I}_3$, set against power density, an appropriate metric for benchmarking different classes of material system; it is also the metric for which estimation protocols are most directly tied to measurable channel parameters (SI S6).

**Relation to the count axis *D*.** Writing $D = log_2 N$ recovers $H_{max}$, so the count axis is the maximum-entropy ceiling of $I_1$. At fine coarse-graining (molecules, proteins, cells) it is approximately $I_1$ itself, which is why the $D$ axis is approximately monotone in observed cross-class data. But because $D$ is a ceiling and not a measure of how that ceiling is occupied, it is silent on $I_2$ and $I_3$: a frozen system and a randomised system with the same $N$ have the same $D$ but very different functional content. The triple ($I_1$, $I_2$, $I_3$) recovers the cross-class ordering of $D$ at zeroth order and resolves the degeneracies that $D$ cannot.

**The three primitives are not algebraically reducible.** $I_1$ is a property of the empirical microstate distribution $p$. $I_2$ is a property of the function-quality landscape $\varphi$ over the design space, independent of $p$. $I_3$ is a property of the channel kernel $p(y|x,\cdot)$, independent of both $p$ and $\varphi$. The three exist on different domains and answer different design questions: *what does the material occupy*, *how rare is the function*, and *how much stimulus information is written into the response*. There is no canonical inner product on the joint space and no information-theoretic justification for a weighted sum $\alpha_1 I_1 + \alpha_2 I_2 + \alpha_3 I_3$. The plane is therefore presented as a family of coordinates, not a scalar collapse.

**What each primitive suggests as design-limiting.** In a *responsive* system the design-limiting primitive is $I_3$: how much stimulus information the material's static channel can transfer, decoupled from how it explores its state space. In an *adaptive* system $I_1$ becomes co-limiting with $I_3$: the internal state $S$ must occupy a sufficient region of state space to support the conditional channel of eqn. (2). In an *intelligent* system all three are co-limiting: $I_3$ for the per-step channel, $I_1$ for the state space supporting the feedback memory, and $I_2$ for the rarity of architectures that maintain feedback stability without losing responsiveness. This is the content of the qualitative observation that intelligence requires more than sensitivity.

## 5. The non-monotonic complexity–function relation and a scaling ceiling

The questions of Section 1 — what bounds $I(X;Y)$, what fraction is realisable, and how both scale with internal complexity — can in principle be answered quantitatively when the channel is the material itself.

**The channel-is-material constraint.** In classical channel-capacity theory the channel is an external object whose properties (bandwidth, noise floor) are specified independently of the message. For a material implementing eqns. (1)–(3) this separation fails: the channel kernel and the physical substrate are the same object. An increase in internal complexity (more coupled molecular degrees of freedom, a deeper cascade of intermediate elements) raises the upper bound on the information that can in principle be transferred, but also introduces additional noise and dissipation through the same elements. The two effects are opposed.

**Linear bound on the upper limit.** For a material with $N$ coupled internal elements, the upper bound on $I(X;Y)$ scales at most linearly in $N$. Each element contributes at most $c$ bits of capacity (set by its local signal-to-noise ratio); cascading or parallelising $N$ such elements with statistically independent noise gives $I_{max} \leq Nc$. This bound is tight only when the elements are statistically independent, which the channel-is-material constraint generally precludes.

**Cascade attenuation.** Consider a cascade of $n$ weakly coupled elements with uniform per-step transmission efficiency $p$ (the fraction of input information passed to the next element). Under the framework of strong data-processing inequalities (SDPI), and as a corollary of the SDPI cascade theorem under uniform contraction [S1],

$$I(X;Y_n) \leq p^n \cdot I(X;Y_0) . \tag{9}$$

The signal energy after $n$ stages is roughly $E_{signal} \cdot p^n$. When this falls below $k_B T$, the downstream element cannot distinguish signal from thermal noise and is decoupled from upstream elements. Setting $E_{signal} \cdot p^{n*} = k_B T$ gives the critical depth

$$N^* \approx ln(E_{stimulus} / k_B T) / ln(1/p) . \tag{10}$$

The heuristic for eqn. (10) is a "signal-drowns-in-noise" argument, which stems from geometric attenuation $E_n = E_{stimulus} \cdot p^n$. Each additional stage in a cascade subtracts a fixed amount, $ln(1/p)$, from the log-energy, so the energy falls by a constant factor per stage. Setting the residual energy equal to the thermal floor, $E_n^* = k_B T$, and solving for the stage index gives eqn. (10). The numerator is the number of "e-folds" of dynamic range between the stimulus energy and the thermal floor; the denominator is how much of that range each stage consumes. Their ratio is the number of stages the signal survives. This is the finite cascade depth set by the stimulus-to-thermal energy ratio and the per-stage transmission $p$, independent of the chemical identity of the elements. Equation (10) maps the SDPI contraction coefficient $p$ (information-theoretic) to the fraction of energy passed per stage; this mapping is assumed, not derived. Beyond $N^*$ further elements add no information transfer but continue to add dissipation. The equality $E_{signal} \cdot p^{n*} = k_B T$ that undepins eqn. (10) holds only under three conditions. (i) *Geometric energy attenuation*: each stage passes a fixed fraction $p$ of the energy it receives, so the signal energy degrades multiplicatively as $E_{signal} \cdot p^n$. This is the weak-coupling, linear-response regime; it fails if a stage has gain, saturates, or is strongly nonlinear. (ii) *A thermal noise floor*: the relevant noise at each element is thermal, of order $k_B T$ per degree of freedom, so an element ceases to transmit when its signal energy reaches that floor. Where the dominant noise is not thermal (shot noise, quantisation, active biochemical noise), $k_B T$ must be replaced by the appropriate noise energy. (iii) *Separability of signal and noise*: the criterion treats "signal energy" as a single scalar that can be compared with a noise energy, which presumes the stages share a common energy currency and that cross-stage correlations are negligible. The equality is therefore an order-of-magnitude decoupling criterion, valid to within a small numerical factor, rather than an exact threshold. Measurements of TNF signalling [24] show end-to-end mutual information of ~1 bit through cascades long enough that this attenuation is mechanistically observable, and quantitatively consistent with eqn. (10) under physiological noise conditions.

**The optimum $N^*$ and a scaling ceiling.** The two effects (linear capacity bound, *rises* with $N$, cascade-survival bound, *falls* with N) together produce a non-monotonic dependence of realised $I(X;Y)$ on $N$:

$$I_{realised}(N) \approx min\{Nc, p^{N-1} \cdot I_{input}\}, \tag{11}$$

whose lower-envelope structure is derived in SI S2. The optimum $N^*$ coincides with $N^*$ of eqn. (10) when the per-element capacity is of order $k_BT$ per stage. This is, in effect, a design rule linking the optimal internal complexity of an adaptive channel to the per-step coupling efficiency, the stimulus magnitude and the operating temperature. It plays a role analogous to the Carnot efficiency in thermal machines (a theoretical maximum that real systems approach but cannot exceed without violating the underlying inequalities) but, unlike Carnot, $I_{realised}(N)$ is a scaling rule, not a theorem, and is not a rigorous equilibrium thermodynamic result.

The position of $N^*$ in the molecular-energy regime relevant to soft matter (per-element transition energies 1–10 $k_BT$) is a substantive conclusion. For $p = 0.5$ and $E_{stimulus} = 5k_BT$ (typical of hydrogen-bond-mediated transitions in aqueous polymers), $N^* \approx ln(5)/ln(2) \approx 2.3$ elements; for $p = 0.9$ and $E_{stimulus} = 10^2 - 10^4 k_BT$ (closer to silicon switching), $N^* \approx 44 - 88$ elements. The order-of-magnitude difference between the soft- and hard-matter regimes is consistent with the architectural divide described empirically in Section 7.

The uniform-$p$ assumption is unlikely to hold strictly for heterogeneous adaptive materials, where successive stages involve different chemistries, relaxation timescales and noise mechanisms. Replacing $p$ with $\{p_i\}$ has four consequences. First, eqn. (9) becomes $I(X;Y_n) \leq (\prod p_i)\, I(X;Y_0)$, and the scalar $N^*$ of eqn. (10) is recovered with $p$ replaced by the geometric mean, but only on average. Second, because contributions enter logarithmically, the smallest $p_i$ dominates the budget: a single stage at $p_i = 0.1$ consumes the depth budget of $\sim$3.3 stages at $p_i = 0.5$ — the origin of the “upstream bottleneck” identified empirically in TNF signalling [24]. Third, if the $\{p_i\}$ are drawn stochastically, the product is approximately log-normal by the central-limit theorem, so practical performance is captured by the median, not the mean. Fourth, $N^*$ becomes a functional of the stage distribution; order-of-magnitude estimates require identifying the rate-limiting stage rather than averaging over stages.

**Thermodynamic coupling.** The energetic cost of operating an adaptive channel is bounded below by Landauer's principle [25], generalised to non-equilibrium steady states by stochastic thermodynamics [26] and assessed for information-processing systems by [27]. Combining the volumetric form of the information rate with the per-volume power gives the volumetric power–rate relation

$$P_{min} = (k_BT\ ln\ 2) \cdot \left(\frac{\dot{I}_3}{V}\right). \tag{12}$$

The underlying steady-state inequality is

$$\sigma \geq k_B (ln\, 2) \cdot \dot{I}_3 \,, \tag{13}$$

where σ is the entropy-production rate. Any non-equilibrium steady-state process maintaining a mutual-information rate $\dot{I}_3$ between two coupled stochastic variables must produce entropy at rate at least $k_B\ ln\ 2 \cdot \dot{I}_3$. This is broader than Landauer's erasure bound; its applicability requires three conditions: (i) the system is at or close to a non-equilibrium steady state; (ii) the mutual-information rate has been operationally identified (input and output have a well-defined joint distribution); (iii) the dissipation σ is causally attributable to establishing or maintaining that coupling. Saturation has been demonstrated experimentally to within a few percent in the quasi-static limit [28]. For an intelligent material with feedback, eqn. (6) modifies the budget: the mutual information acquired through feedback reduces the minimum work for a given state change by up to $k_BT$ (ln 2) per cycle.

Equations (12) and (13) together give the maximum stimulus-to-response information rate achievable at internal complexity $N^*$, per unit volume, at operating temperature, bounded above by the volumetric power dissipation divided by $k_BT$ ln 2. This is the floor (benchmark) against which all classes of stimulus-coupled matter can be compared.

**What the ceiling implies for design.** Equations (10)–(13) imply three design guidelines for adaptive soft matter that are not generally evident in the literature.

*(i) There is a finite optimal internal complexity.* Adding responsive elements beyond $N^*$ does not necessarily improve function and incurs dissipation. The widespread assumption that more chemistry (more components, more reaction steps) is better is not supported by the channel analysis.

*(ii) The optimal complexity depends on stimulus magnitude.* A system designed for a strong stimulus ($E_{stimulus} \gg k_BT$) tolerates a deeper cascade than one for a weak stimulus. The same architecture is not optimal for both regimes.

*(iii) Feedback changes the energy budget.* An intelligent material exploits eqn. (6) and operates closer to the Landauer–Bérut benchmark than a non-feedback adaptive material with the same $I_3$. Converting responsive to intelligent function therefore yields both a thermodynamic and a functional gain, at the cost of the architectural complexity required to implement the feedback path.

The dimensional matching of $\frac{\dot{I}_3}{V}$ to a volumetric power density gives a single thermodynamically anchored plane on which any stimulus-coupled material can be located. The irreducibility of $I_1$ and $I_2$ to $I_3$ requires a companion static plane to retain the structural information that the dynamic plane discards.

## 6. The two benchmarking planes

The reference class for "hard" systems comprises digital silicon (CMOS SoC, GPU), neuromorphic analogue arrays (memristor crossbars) and electromechanical actuators (MEMS, piezoelectric, shape-memory alloy). Each subclass highlights a different relationship between the three primitives and the volumetric power density, and the asymmetries among them are themselves design information.

**Plane A — dynamic ($\dot{I}_3/V$, P).** Power density $P$ (W·m$^{-3}$) is introduced as the abscissa for three reasons. First, it is the natural *cost* conjugate to the information *rate*: the Landauer relation (eqn. 12) makes $P$ and $\frac{\dot{I}_3}{V}$ dimensionally matched through the single constant $k_B T \ln 2$, so the two axes can be plotted together and their ratio is a pure number. Second, that ratio $P/P_{min}$ is what discriminates between material classes: two systems may transduce information at the same volumetric rate yet sit decades apart in $P$, and it is the vertical distance above the Landauer diagonal — the dissipative overhead per bit — that separates evolved biology (close to the benchmark) from silicon (far above it) and locates where a candidate adaptive material falls between them. Third, $P$ is an absolute, instrument-measurable quantity (calorimetry, electrical power, metabolic flux) that does not depend on the coarse-graining and reference conventions that limit $I_1$ and $I_2$; it therefore anchors the otherwise convention-laden information metrics to a hard thermodynamic scale. Volumetric normalisation (per m$^3$) is used so that systems spanning a femtolitre organelle to a cubic-centimetre gel can be compared on the same axes.

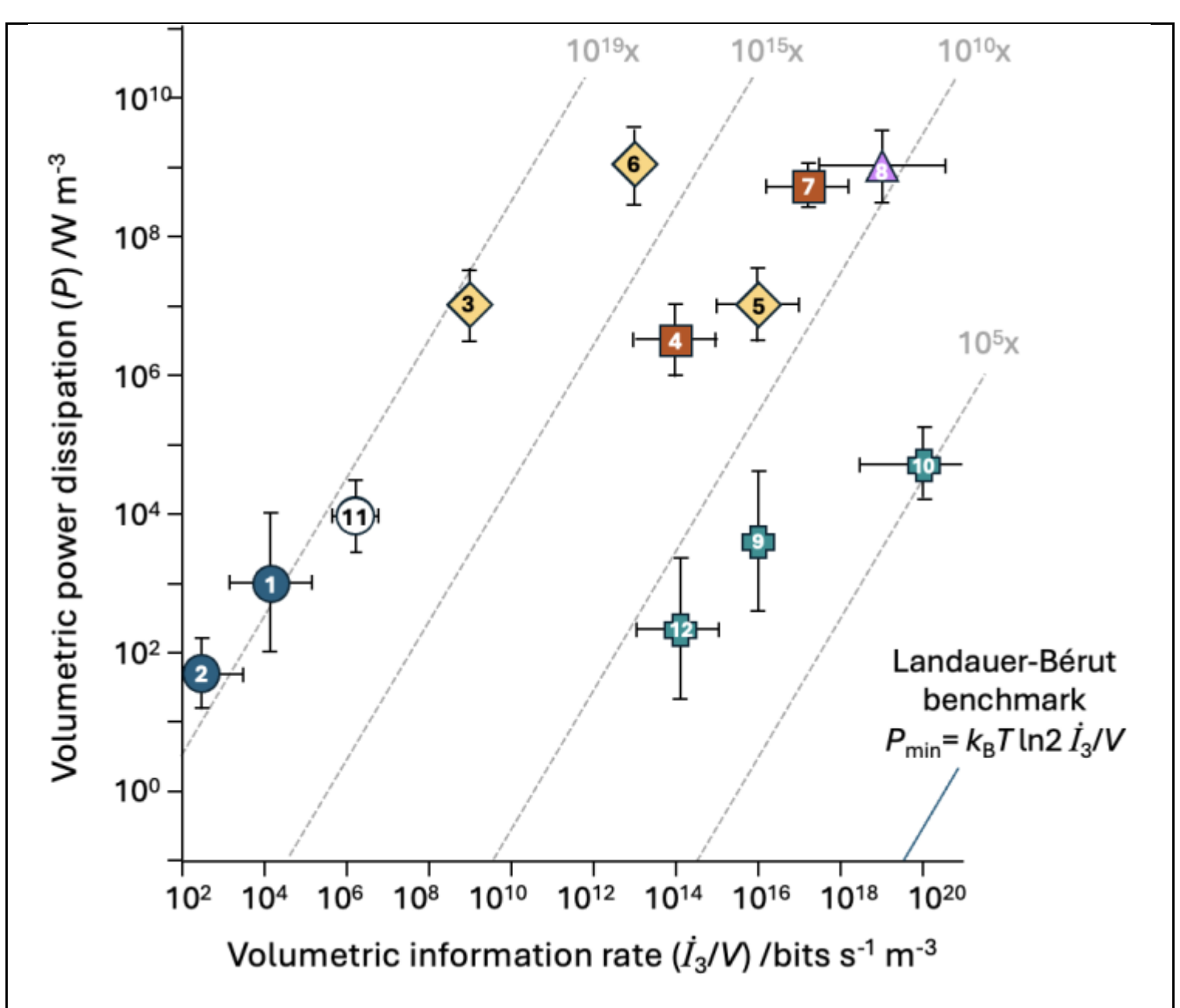

**Figure 2.** Plane A (I₃/V, P). 1: PNIPAM hydrogel, 2: Glucose matrix, 3: Nitinol SMA, 4: BLE SoC (active), 5: MEMS accelerometer, 6: piezoelectric stack, 7: GPU die, 8: Memristor crossbar, 9: *E. coli* chemotaxis, 10: mitochondrion, 11: BZ-PNIPAM gel [S8], 12: ER membrane system [S9].

The volumetric information-rate axis $\frac{\dot{I}_3}{V}$ allows three constructions depending on what the system is doing: (i) the Shannon stimulus–response rate, (ii) the computational bit-manipulation rate, and (iii) the cyclic-response rate. These can coincide in special cases (an analogue-to-digital converter at its channel capacity) but in general they differ by orders of magnitude. The Plane A diagram (Fig. 2) benchmarks each system against its own Landauer benchmark, computed from whichever rate is appropriate to its function. Cross-class comparison using the generic $\dot{I}_3$ notation indicates dissipation efficiency of bit manipulation broadly construed, not efficiency of stimulus–response transduction specifically. The plane is logarithmic in both axes: the abscissa is $P$ ($W \cdot m^{-3}$), the volumetric power dissipation under operating conditions; the ordinate is $\frac{\dot{I}_3}{V}$ ($bits \cdot s^{-1} \cdot m^{-3}$), the volumetric stimulus-to-response information rate. The Landauer–Bérut benchmarking scale [25,28] is the diagonal eqn. (12),

$$P_{min} = (k_B T \ ln\ 2) \cdot (\frac{\dot{I}_3}{V}),$$

a single straight line of unit slope. Every adaptive or intelligent system must lie on or above it. The multiplicative excess $P/P_{min}$ is the system's single dimensionless figure of merit, comparing the operating efficiency of any stimulus-coupled material to the thermodynamic limit set by the bit rate it processes. For adaptive and intelligent systems, where internal state updates carry erasure content and feedback brings the Sagawa–Ueda relation (eqn. 6) into play, the diagonal is a genuine thermodynamic floor on the dissipation causally required to maintain the input–output coupling. For purely responsive systems such as PNIPAM, by contrast, the dissipation is dominated by molecular hysteresis (hydrogen-bond rearrangement, viscoelastic loss, chain disentanglement) and has no causal connection to the information transduced; the inequality is trivially satisfied and the diagonal functions not as a floor but as a $k_B T$-per-bit reference scale against which the thermodynamic overhead of non-informational dissipation can be quantified. The Landauer ratio $P/P_{min}$ should be read accordingly: a binding efficiency metric for the adaptive and intelligent classes, and a benchmarking scale otherwise.

**Plane B — static ($I_1$, $I_2$).** The companion plane to Plane A is linear in $I_1$ (configurational diversity at operational coarse-graining) and logarithmic in $I_2$ (Hazen functional selectivity). It has no fundamental reference line: it is descriptive, and exists because the dynamic plane discards the structural information that distinguishes a generic responsive medium from an evolved or precision-engineered system at the same coordinates. A point on Plane A is interpreted via its companion point on Plane B. Both axes of Plane B are convention-dependent — $I_1$ through the coarse-graining choice and $I_2$ through the reference design space and threshold — so comparisons on Plane B carry larger uncertainty than those on Plane A, and absolute placements should be treated as order-of-magnitude at best (§6.3). Making the two axes more robust requires community-agreed conventions for coarse-graining ($I_1$) and for the reference chemistry space and functional threshold ($I_2$).

### 6.3 Scope and limitations of the framework

The framework compares thermodynamic efficiencies of information processing, not the semantic value of the information processed. A GPU performing $10^{14}$ bit-operations and a bacterium transducing 1 bit of chemotaxis information sit at the same Landauer benchmark *per bit*; the framework makes no claim that the two systems are functionally equivalent.

Two limitations are explicit. First, $I_3$ is defined relative to an identified stimulus–response channel. For systems whose primary natural function is internal energy conversion (mitochondria, Li-ion cells), the metabolic-flux “channel” is not stimulus-coupled, and the framework's $I_3$ is computed only for the regulatory channel, where it remains well defined; in the mitochondrial case $I_3$ refers to stimulus–response mutual information at the organelle's regulatory interface, not to internal metabolic-flux information. Second, all numerical placements carry uncertainty of typically ±1–2 decades on each axis; the qualitative ordering is robust but the absolute placements are not. $I_2$ is the weakest of the three primitives: useful as a qualitative axis distinguishing precision-engineered from generic systems, but absolute placements should carry explicit conventions and be treated as order-of-magnitude.

## 7. Application of the framework

Sixteen system spanning hard materials through synthetic soft matter to biological systems have been selected to illustrate how the framework is applied. The protocols used to estimate the three primitives are described in the Supplementary Information. The PNIPAM hydrogel, glucose-responsive matrix and *E. coli* chemotaxis pathway recapitulate the responsive, adaptive and intelligent classes of Section 2, with *E. coli* the only system whose stimulus-to-response rate has been directly measured. The eukaryotic endoplasmic-reticulum (ER) lipid-biosynthesis system is included as a functional system that achieves robustness by integrating multiple inputs into a single feedback signal (worked up in full in SI S9). A mitochondrion is included as a biological organelle-scale reference, with the understanding that (i) its primary function is internally regulated energy conversion rather than external stimulus response and (ii) it carries the greatest placement uncertainty; its Plane A positioning uses a regulatory-pathway information rate as a proxy for $\dot{I}_3$, so is only an order-of-magnitude estimate. The information primitives of four further soft-matter systems reported in the literature [5] [7][38] [39], together with two commercial soft matter therapeutic systems, are estimated to illustrate the discriminatory power of the framework. The six hard systems are estimated from peer-reviewed sources under the same uncertainty conventions. $T$ = 298 K is assumed for synthetic systems and 310 K for in vivo biological and implant systems; the difference (~4 %) is immaterial at the scale of the plane.

**Soft-matter systems, synthetic.** The PNIPAM hydrogel and the glucose-responsive insulin-release matrix recapitulate the responsive and adaptive classes of Section 2. Both operate by reorganising weak molecular interactions — a few $kJ \cdot mol^{-1}$ per repeat unit [8] — over slow cycles, so each transduces only a small number of bits per excursion and dissipates a substantial fraction of its reorganisation enthalpy as heat. The result is a low volumetric information rate and a very large dissipative overhead, placing both deep in the soft-matter band high above the Landauer floor. The PNIPAM gel is purely responsive (a binary thermal cue mapped to a graded swelling state); the glucose matrix is adaptive, its phenylboronate loading state conditioning the release response to recent glycaemic history. Numerical placements for both systems are collected in Table 1, with the underlying $I_1$, $I_2$ and $\frac{\dot{I}_3}{V}$ estimates and their uncertainties derived in SI S4–S7.

Four further synthetic soft-matter systems drawn from the recent literature are placed alongside these to test the discriminatory power of the framework. All four are responsive under the strict kernel criterion of Section 2 (i.e. none closes an output-to-input regulatory loop) yet they differ markedly in architecture, and the two planes separate them where a single complexity scalar could not. Their numerical placements are collected in Table 1 and their qualitative positioning in Table 2; the underlying calculations are given in SI S11–S14.

*Gabrielli et al. responsive router* [38]. A bulk-solution system maps two nucleotide inputs (GMP, TMP) onto two surfactant-assembly outputs by selective templating, with the response made transient by enzymatic (alkaline phosphatase) trigger cleavage. It is a genuine 2→2 demultiplexing kernel with signal-transduction-like transience, but the mapping does not update, there is no feedback path, and the reduced second-cycle response is depletion rather than adaptation. It is therefore responsive, at the information-processing end of that band. The noisy channel transfers $I_3$ ≈ 0.2 bit per cycle at $\nu \approx 3 \times 10^{-4}$ s$^{-1}$; on a ~1 mL reaction volume this gives $\dot{I}_3/V \approx 6 \times 10^{1}$ bits·s$^{-1}$·m$^{-3}$ and, from enzymatic phosphate hydrolysis, $P \approx 0.2$ W·m$^{-3}$, placing it in the lower-left of the soft-matter band at $P/P_{\min} \approx 10^{18}$ (SI S11).

*Adamska et al. multistate assembly* [5]. A single isoleucine–naphthalenediimide component is steered into five topologically distinct chiral assemblies by temperature, solvent, concentration and guest, including $C_{60}/C_{70}$ shape-discrimination. The state multiplicity exceeds Gabrielli's, but the system is explicitly thermodynamically controlled: every state is an equilibrium selected by externally imposed conditions, with no dissipative engine, fuel or intrinsic clock. It is therefore multistate responsive, and it cannot be placed on Plane A. This is because at equilibrium the power density vanishes and the information rate is set by the operator, not the material, so any ($\dot{I}_3$, $P$) point would sit below the Landauer floor, which is forbidden. Its information is structural and belongs on Plane B, where it scores $I_1$ ≈ 2–2.5 bits over the five states and a high functional selectivity $I_2$ ≈ 30–40 bits (SI S12).

*Chiolerio et al. polyaniline colloid* [7]. A sub-percolation PANI-nanorod colloid driven by a 10 V AC field undergoes a ~2-decade resistance change that persists after the field is removed. This is a non-volatile, history-integrating state that is the memory primitive both preceding systems lack. It is the furthest toward adaptive of the four on the memory axis, but it remains a single plastic element with no closed loop, hence responsive-with-plasticity. Being actively driven and dissipative, it is placeable on Plane A: $I_3$ ≈ 1 bit per write at $\nu \approx 2 \times 10^{-2}$ s$^{-1}$ gives $\dot{I}_3/V \approx 1 \times 10^{4}$ bits·s$^{-1}$·m$^{-3}$, with $P \approx 10^{2}$ W·m$^{-3}$ during stimulation and $P/P_{\min} \approx 3 \times 10^{18}$ (SI S13). Its function is generic, so it scores lowest of the four on the static plane, $I_2$ ≈ 10–15 bits.

*Dúzs et al. mechano-adaptive meta-gel* [39]. The most architecturally complete of the four: a metamaterial strain-gate (sensor) with an all-or-nothing threshold feeds an autocatalytic urea–urease reaction-diffusion front (processor, with gain and long-range transmission) that drives global stiffening or actuation (actuator), this constituting a full sensor-processor-actuator chain. It is nonetheless open-loop (the autocatalysis is feedforward amplification, not regulatory feedback) and single-use (nanofibrillation is an irreversible write), so it stops short of closed-loop adaptation. Fuel-driven and non-equilibrium, it is placeable on Plane A, where a one-shot ~1-bit hour-scale decision scores modestly: $\dot{I}_3/V \approx 3 \times 10^3$ bits·s$^{-1}$·m$^{-3}$, $P \approx 4 \times 10^2$ W·m$^{-3}$, $P/P_{\min} \approx 3 \times 10^{19}$ (SI S14). Its sophistication is carried instead by the gain and by a high static selectivity $I_2 \approx 40$–$60$ bits.

Taken together the four make the framework's central methodological point concretely: on Plane A the three placeable systems sit within ~2 decades of one another in the responsive band ($10^{18}$–$10^{19}$ above the floor) and the fourth cannot be placed at all, so the dynamic coordinates separate them only weakly. What actually distinguishes them, equilibrium versus driven (Adamska), non-volatile memory (Chiolerio), processor/gain architecture (Dúzs), engineered recognition (Gabrielli) and topological selectivity (Adamska), is carried by the kernel structure and by Plane B ($I_2$), not by ($\dot{I}_3/V$, $P$). Given the  limitations of $I_2$ the relative placements of the four systems make heuristic sense but require validation. However, the orthogonality of the memory axis and the static-selectivity axis highlights why the framework keeps the two planes separate rather than collapsing them onto a single complexity scalar.

**Soft-matter systems, biology.** Three biological systems anchor the evolved end of the continuum. A single *mitochondrion* is included as an organelle-scale reference; because its primary function is internally regulated energy conversion rather than external stimulus response, its Plane A placement uses the information rate of its regulatory interface as a proxy and carries the largest uncertainty of any entry. *E. coli* chemotaxis is the one system whose stimulus-to-response rate has been measured directly [30]; its pathway power budget is reconstructed from primary measurements of receptor methylation [31], CheA/CheY phosphotransfer [32] and flagellar-motor dissipation [33] (SI S7). The *eukaryotic endoplasmic-reticulum (ER) lipid-biosynthetic network* [34] is a functional system that maintains homeostasis by integrating many lipid inputs into a single slow feedback signal (worked up in full in SI S9).

The descriptive picture from Table 1 is that evolved systems sit closest to the Landauer floor, but not uniformly: the mitochondrion has lower per-bit thermodynamic overhead than the ***E. coli*** chemotaxis pathway, even though its absolute power density is higher. The chemotaxis pathway therefore sits one to three decades further above the floor despite dissipating less power. The ER network sits between the soft-matter band and the fast-biology band, at high $I_2$ on the static plane, because its integrative-feedback architecture buys robustness rather than information-rate efficiency. The self-consistent set of values used here for the ER system is derived in SI S9, but it is noted that they are obtained by extrapolation from a non-information-theoretic source. All numerical placements and uncertainties for the three systems are given in Table 1.

**Hard-matter systems, digital.** The Bluetooth-Low-Energy SoC and the data-centre GPU represent digital silicon. Their per-element switching energy is ~$10^4$–$10^5$ $k_B T$ [35], three to four decades above the soft-matter molecular scale, so they achieve high volumetric information rates only at correspondingly high power densities. The GPU has the highest volumetric information rate of any system on the plane below biology, but also the highest power density; the BLE SoC trades rate for a much lower duty-cycle-averaged dissipation. For these systems $I_1$ and $I_2$ are estimated from operational state-count and design-space cardinality respectively (SI S4–S5) rather than derived, and the placements are tabulated in Table 1.

**Hard-matter systems, neuromorphic.** The memristor crossbar performing in-memory analogue inference is the engineered system that comes closest to the Landauer floor. By suppressing the data movement that dominates digital architectures [36], it sits about two decades below the GPU in dissipative overhead at the same task, leaving a remaining gap to biology of roughly five decades. The improvement is architectural — it comes from eliminating data transport, not from reducing the per-device switching energy. Placement values are given in Table 1; the estimation conventions are those of SI S4–S7.

**Hard-matter systems, electromechanical.** The MEMS accelerometer, the piezoelectric stack actuator and the Nitinol shape-memory-alloy (SMA) actuator span the electromechanical class. The accelerometer is a high-bandwidth capacitive sensor; the piezoelectric and SMA actuators are binary-to-graded transducers operating at much higher per-element energies than soft matter. The SMA actuator is the hard system that sits closest to soft matter on Plane A [37] — a quantitative reflection of the fact that its thermally driven martensitic transformation has a per-element thermodynamic structure not unlike a polymer phase transition. Numerical placements for all three are collected in Table 1, with derivations in SI S4–S7.

| System | $I_1$ (bits) | $I_2$ (bits) | $\dot{I}_3/V$ (bits·s$^{-1}$·m$^{-3}$) | P (W·m$^{-3}$) | $P/P_{min}$ |
|---|---|---|---|---|---|
| *PNIPAM hydrogel* | 1.6 ± 1 | 13 ± 4 | $1.5 \times 10^{4}$ | $10^{3}$ | $2 \times 10^{19}$ |
| *BZ-PNIPAM hydrogel* | 3 | 27 ± 4 | $10^{6}$ | $10^{4}$ | $3.6 \times 10^{18}$ |
| *Glucose-responsive matrix* | 2 ± 1 | 20 ± 3 | $2.5 \times 10^{2}$ | 50 | $7 \times 10^{19}$ |
| *Gabrielli responsive router* | ~1 | ~28 | $6 \times 10^{1}$ | 0.2 | ~$10^{18}$ |
| *Adamska multistate assembly* | 2–2.5 | 30–40 | — | — | — |
| *Chiolerio PANI colloid* | 1–2 | 10–15 | $1 \times 10^{4}$ | $10^{2}$ | $3 \times 10^{18}$ |
| *Dúzs mechano-adaptive meta-gel* | 1–2 | 40–60 | $3 \times 10^{3}$ | $4 \times 10^{2}$ | $3 \times 10^{19}$ |
| *Nitinol SMA actuator* | 1 ± 0.5 | 30 ± 5 | $10^{9}$ | $10^{7}$ | $3 \times 10^{18}$ |
| *BLE SoC (active)* | ~5 | ~80 | $10^{14}$ | $3 \times 10^{6}$ | $10^{13}$ |
| *MEMS accelerometer* | ~3 | ~50 | $10^{16}$ | $10^{7}$ | $3 \times 10^{11}$ |
| *Piezoelectric stack actuator* | 1 ± 0.5 | 30 ± 5 | $10^{13}$ | $10^{9}$ | $10^{16}$ |
| *GPU die* | ~10 | ~$10^{2}$ | $1.7 \times 10^{17}$ | $5 \times 10^{8}$ | $10^{12}$ |

| System | $I_1$ (bits) | $I_2$ (bits) | $\dot{I}_3/V$ (bits·s$^{-1}$·m$^{-3}$) | P (W·m$^{-3}$) | $P/P_{min}$ |
|---|---|---|---|---|---|
| *Memristor crossbar* | ~5 | ~80 | $10^{19}$ | $10^9$ | $10^9$–$10^{10}$ |
| *E. coli chemotaxis* | 1–2 | $10^3$–$10^4$ | $10^{16}$ | $4 \times 10^3$ | $10^8$ |
| *Mitochondrion (interface)* | 3 ± 1 | ~$10^5$ | $10^{20}$ | $5 \times 10^4$ | $1.7 \times 10^5$ |
| *Eukaryotic ER membrane* | 2.5–4 | ~$10^4$ | $5 \times 10^{13}$ | $10^4$–$10^5$ | ~$10^{11}$ |

**Table 1.** Order-of-magnitude estimates of the three information-theoretic metrics and the power density for the sixteen systems. Italicised ER row recomputed for internal consistency (see §7.2). The Adamska multistate assembly is at thermodynamic equilibrium and has no defined Plane A coordinates ($\dot{I}_3/V$, $P$, $P/P_{min}$) marked; it is placed on the static plane ($I_1$, $I_2$) only.

### 7.6 Two industrially relevant systems positioned qualitatively

Two systems of current therapeutic importance — the Camurus FluidCrystal® injectable depot and the BioNTech BNT162b2 mRNA–lipid-nanoparticle (LNP) vaccine — are positioned qualitatively in the main text rather than plotted in Fig. 2, because neither admits a *convention-free* stimulus–response $\dot{I}_3$ or power density: the choice of functional unit and operating timescale moves the result by more than fifteen decades. Convention-dependent order-of-magnitude estimates are nonetheless instructive, and are worked in full in SI S10; the resulting values are quoted below with their interpretation. Both systems are included because they show how the channel taxonomy applies to real, deployed soft-matter products.

**FluidCrystal depot.** The injected lipid solution (glycerol dioleate / soy phosphatidylcholine with trace *N*-methyl-2-pyrrolidone) [40] self-assembles *in situ* as water diffuses in, transforming through a sequence of lyotropic phases (a reverse-hexagonal outer shell over a reverse-micellar-cubic interior) that sets the diffusional release rate. The release kernel is not time-invariant: it evolves with accumulated hydration, and hydration-driven lipid redistribution establishes persistent composition and structure gradients that govern release over weeks. In the channel formalism the current map therefore depends on an internal state set by input history, $p(y|x,s), s = s(X_{t-1})$ — the signature of adaptive matter. What keeps it short of a definite adaptive system is the character of that state: hydration is a single, monotone, saturating stimulus, so the internal state relaxes one-way toward equilibrium rather than being reversibly re-conditioned by a varying physiological signal, and there is neither output-to-channel feedback nor multi-signal integration. The system therefore sits at the responsive–adaptive border, on the responsive side. Treating the whole depot as the functional unit (SI S10), the convention-dependent estimates are $\frac{\dot{I}_3}{V} \approx 10^0$–$10^1$ bits·s$^{-1}$·m$^{-3}$, $P \approx 10^1$–$10^2$ W·m$^{-3}$ and a Landauer ratio $P/P_{min} \approx 10^{21}$–$10^{22}$. This is the lowest volumetric information rate and the highest dissipative overhead of any system considered — the limiting case of a near single-shot delivery device that transduces almost no information while dissipating real release power.

**BNT162b2 mRNA-LNP.** Material and biological function must be assessed separately. At the material level the four-lipid LNP (ALC-0315, DSPC, cholesterol, ALC-0159 at 46.3 : 9.4 : 42.7 : 1.6) [41] is a clean responsive system: the ionizable lipid (apparent $pK_a \approx 6.0–6.5$) is neutral at administration and protonates in the acidic endosome, a single-cue map $p(y|x)$ executed once — pH in, endosomal escape out — after which the carrier is spent. It has *less* material-intrinsic adaptive character than FluidCrystal, because there is no slow, persistent, release-conditioning state. At the *system* level, however, the mRNA program is translated into antigen that engages clonal selection, immunological memory and germinal-centre affinity maturation — a history-dependent, feedback-modified biological channel $p_{null}(y_t|x_t, X_{t-1}, Y_{t-1})$ of exactly the intelligent form, sitting orders of magnitude closer to the thermodynamic floor than any synthetic soft matter. The decisive point is structural: FluidCrystal must become adaptive within the material, whereas BNT162b2 borrows adaptive/intelligent behaviour by interfacing with a pre-existing biological adaptive system. Treating the single nanoparticle as the functional unit during its endosomal-escape window (SI S10), the material-level estimates are $\frac{\dot{I}_3}{V} \approx 10^{18}–10^{19}$ bits·s$^{-1}$·m$^{-3}$ ($\approx 10^{17}$ when weighted by the 1–2 % escape efficiency), $P \approx 10^3–10^4$ W·m$^{-3}$ and a Landauer ratio $P/P_{min} \approx 10^5–10^7$ — a biology-like value, because the escape event is a small, fast, near-single-bit molecular switch. The system-level immune channel is not assignable to the material. The two products therefore sit at opposite extremes of Plane A, and the ≈16-decade gap between them is set entirely by the unit of analysis — a bulk depot acting over weeks versus a single nanoparticle acting over minutes — not by any difference in the underlying lipid physics (SI S10).

| System | Class (material) | Class (system) | Internal-state memory | Closest route up |
|---|---|---|---|---|
| Glucose matrix | Adaptive | Adaptive | State-conditioned (swelling/loading) | Add output feedback |
| BZ-PNIPAM gel | Adaptive (2 of 3 intelligence routes) | Adaptive | BZ chemical state (oscillator) | Add output-to-channel feedback |
| Gabrielli responsive router | Responsive (2→2 transduction) | Responsive | None (resets each cycle) | Add output→input feedback |
| Adamska multistate assembly | Responsive (multi-state, equilibrium) | Responsive | None (equilibrium) | Add dissipative drive + feedback |
| Chiolerio PANI colloid | Responsive + plasticity | Responsive | Non-volatile (conductance) | Close the regulatory loop |
| Dúzs mechano-adaptive meta-gel | Responsive + processor/gain | Responsive | Irreversible write (one-shot) | Add negative feedback / resettability |

| System | Class (material) | Class (system) | Internal-state memory | Closest route up |
|---|---|---|---|---|
| BNT162b2 mRNA-LNP | Responsive (clean one-shot) | Adaptive/intelligent (borrowed) | None at material level | Program adaptivity into cargo |
| ER lipid network | Adaptive–intelligent | Intelligent (integrative feedback) | Stored elastic energy (torque) | — (reference biology) |
| E. coli chemotaxis | Intelligent | Intelligent | Methylation memory | — (reference biology) |

**Table 2.** Qualitative channel-class positioning of the worked example systems of §7.1–7.6. Here (reference biology) denotes a naturally evolved system included as a benchmark against which the synthetic systems are compared, rather than as a design target to be improved. Because such systems already operate close to the thermodynamic floor and embody fully developed adaptive or intelligent feedback, there is no "route up" to specify for them; they instead mark the position that engineered soft matter would have to approach. The Closest route up column is therefore left blank for the ER lipid network and *E. coli* chemotaxis.

### 8. Mechanistic interpretation of the Landauer-ratio clustering

Within the limitations of the framework's assumptions and of the estimates of key quantities, the systems fall into broad bands on the dynamic plane. Comparison across bands indicates how far each system sits above its own Landauer benchmark, but band membership does not directly translate into stimulus-response efficiency. For example, hard digital systems dissipate $\sim 10^{10}\times$ Landauer to perform internal computation, while *E. coli* dissipates $\sim 10^{8}\times$ Landauer to transduce an external stimulus; these are not the same kind of efficiency.

*Band 1 — soft matter and SMA: $10^{18}$–$10^{20}\times$ Landauer*. PNIPAM, the glucose matrix and the Nitinol actuator cluster within one decade despite very different chemistry. Any responsive event involving a 1–10 kJ·mol$^{-1}$ molecular reorganisation per repeat unit, repeated over $\sim 10^{22}$–$10^{23}$ units per cubic centimetre, dissipates a substantial fraction (typically 10–30 %) of the reorganisation enthalpy as heat per cycle. The Landauer ratio is set by the molecular-energy scale per element multiplied by the volumetric density of responding units, divided by the bit rate per element; for all soft-matter and thermal-actuator systems this collapses to roughly $10^{19}$. This is a generic result for molecular condensed-phase responsiveness.

*Band 2 — MEMS, BLE digital, piezoelectric, GPU: $10^{10}$–$10^{16}\times$ Landauer.* These share a silicon-substrate switching architecture. The per-element switching energy is $\sim 10^{-14}$–$10^{-13}$ J at modern CMOS nodes [35], i.e. $\sim 10^{4}$–$10^{5}$ $k_B T$ per switching event — three to four decades above the soft-matter per-element scale. The higher overall plane positions reflect operational overhead (clock distribution, leakage, data movement) that dominates rest-of-system dissipation.

*Band 3 — memristor crossbar: $10^9$–$10^{10}$× Landauer.* In-memory analogue computing eliminates the data-movement overhead that dominates digital architectures [36]. The two-decade improvement over GPUs at the same task brings neuromorphic hardware closer to the floor than any other engineered system on Plane A; the remaining gap to biology is approximately five decades.

*Band 4 — E. coli and the mitochondrion: $10^5$–$10^8$× Landauer.* Evolved cellular and organelle systems sit closest to the floor, but the band is wider than the soft-matter and silicon bands. The mitochondrion (≈ 1.7 × $10^5$) is closer to the floor than the *E. coli* chemotaxis pathway (≈ $10^8$): per bit of stimulus information actually transduced, the mitochondrion has lower thermodynamic overhead, even though its absolute volumetric power dissipation is higher. This is consistent with the cell-level estimates of Parrondo, Horowitz and Sagawa [27] and the chemotaxis-rate measurements of Mattingly *et al.* [30].

*Intermediate point — the ER membrane system:≈ $10^{11}$× Landauer.* The ER lipid-biosynthesis system does not belong to any of the four bands; it sits between Band 1 and Band 4 at ∼$10^{11}$ (range $10^{10}$–$10^{12}$). The mechanism is architectural, not energetic. The ER uses an *integrative, slow* feedback loop: many lipid channels are summed into a single scalar (the membrane torque parameter) that updates on a physiological-perturbation timescale (∼$10^{-3}$ s$^{-1}$). The information rate per unit volume is therefore low, the Landauer floor is correspondingly low, and the ratio $P/P_{min}$ is large despite no individual reaction being thermodynamically wasteful. The ER thus demonstrates that an integrative-feedback architecture can *raise* the Landauer ratio at high $I_2$, because integration over many channels is a route to robustness, not to dissipative efficiency.

## 9. Soft–hard matter differences in the molecular-energy regime

The indicative clustering on Plane A reflects a structural divide in the molecular energetics of the underlying substrate, with consequences for $I_2$, fatigue and information-rate ceilings that follow from the scaling ceiling.

**Per-element energy scales.** The per-element transition or switching energy distinguishes the substrates by three to four orders of magnitude:

• *Soft matter*: 1–10 $k_BT$ per repeat unit (≈ 2.5–25 zJ at 298 K), set by hydrogen bonding, hydrophobic interactions and conformational reorganisation — the regime in which thermal fluctuations are comparable to the driving energy of the responsive transition.

• *SMA, piezoelectric, MEMS*: $10^2$–$10^4$ $k_BT$ per element ($10^{-19}$–$10^{-17}$ J), set by martensitic phase transformations, piezoelectric coupling or capacitive switching.

• *CMOS at modern nodes*: ∼$10^4$–$10^5$ $k_BT$ per switching event (≈ 0.1–1 pJ) [35].

• *Memristor*: $10^2$–$10^3$ $k_BT$ per synaptic operation; closer to the molecular regime than CMOS digital, which is why its Landauer ratio is lower.

• B*iology*: ATP hydrolysis at ~20 $k_BT$ per molecule — comparable per-element scale to soft matter, but used in cycles that achieve much higher conversion efficiency through evolved coupling.

**Consequences for $I_2$.** Functional selectivity $I_2$ tracks the per-element energy scale, but inversely to a naive expectation. Soft-matter responsive transitions occupy a wide region of chemical design space (many polymers exhibit LCST behaviour; many boronate chemistries bind diols), so $I_2$ is low (10–30 bits, on heuristic arguments). Hard-matter systems operate in a narrow region of materials space defined by silicon physics and require precise lithography and dopant placement, so $I_2$ is moderately high (50–100 bits). Biology operates in a narrow region of *sequence* space defined by catalytic and regulatory specificity, and $I_2$ is very high ($10^3$–$10^5$ bits). The ER system illustrates a biological system with a high Landauer ratio at high $I_2$: Beard *et al.* [34] provide a direct numerical demonstration that the sensitivity of the torque parameter to individual rate constants collapses by ~10× when feedback is added at the CCT-catalysed reaction, and further when applied at multiple points, while the sensitivity of individual lipid concentrations is *unaffected* by feedback. This is empirical evidence that the controlled variable acquires high functional selectivity (high effective $I_2$) while the underlying state variables retain low $I_2$.

Two implications follow. First, soft adaptive matter operates at low $I_2$ because chemical responsiveness is common; achieving the selectivity required for intelligent function (orthogonal multi-channel response, decision logic) is a problem of raising $I_2$ without losing $I_3$. Second, evolved biology shows that high $I_2$ and low Landauer ratio are compatible — the rarity of functional configurations and the thermodynamic efficiency of operation are not in tension, contrary to the intuition that “more complex” must mean “more dissipative.”

**Consequences for fatigue.** The 1–10 $k_BT$ per-element energy scale means the driving energy of the responsive transition is comparable to thermal fluctuations. This is the molecular origin of the fatigue problem: repeated cycling reorganises the network slightly each time, because the energy barriers between conformational states are crossed many times per cycle by thermal motion, and a small but cumulative fraction of those crossings produces irreversible changes (covalent scission, crosslink rearrangement, segregation). Hard-matter systems, with per-element energies $10^3$–$10^4$× thermal, do not fatigue by this mechanism; their failure modes (electromigration, dielectric breakdown, mechanical fracture) differ in physics and timescale. The fatigue problem is therefore intrinsic to the energy regime of soft matter and not separable from the responsiveness mechanism.

**The $N^*$ divide.** From Eqn. (10), or the soft-matter regime (per-element energy ~5 $k_BT$, $p \sim 0.5$), $N^*$ ≈ 2.3; for the CMOS regime (per-element energy $10^2$–$10^4$ $k_BT$, $p \sim 0.9$ due to higher fidelity), $N^*$ ≈ 44–88. The order-of-magnitude difference explains why soft-matter architectures cannot be made arbitrarily complex without dissipative collapse, whereas hard-matter architectures scale to billions of elements: the constraint is set by the substrate, not by engineering effort. This separation is relatively insensitive to heterogeneity — a decade-scale variation among $p_i$ does not close a decade-scale gap — though placing any individual system on Plane A requires identifying the cascade bottleneck rather than a global $p$.

The implication for adaptive soft matter is that populating the gap between Band 1 and the biology band requires either (i) raising the per-element energy scale, or (ii) using a feedback architecture in which the thermodynamic budget is shifted by the Sagawa–Ueda relation (eqn. 6). The ER membrane system further illustrates that an integrative-feedback route can *raise* the Landauer ratio at high $I_2$, not only lower it, because integration over many channels buys robustness rather than efficiency.

## 10. A putative design space for adaptive and intelligent soft matter

Between Band 1 (soft, $\sim10^{19}\times$ Landauer) and Band 4 (biology, $\sim10^{8}\times$ Landauer) lies a region of $\sim10^{11}$ in Landauer ratio that is empty of synthetic systems. Populating it requires more than chemical novelty: the framework shows that the Landauer ratio of any soft-matter system using 1–10 $k_BT$ per element is set by molecular energetics, not by the specific chemistry. Populating the gap requires either (a) substantially higher per-element energies, transitioning the substrate toward Band 2, or (b) higher information rates per unit dissipation, achievable through feedback (intelligence) or through parallelisation of channels (multi-channel orthogonality). The three worked biological/synthetic examples of §7 illustrate the available routes.

Three architectural elements could position soft matter within this design space.

*(i) A feedback path from output to channel.* This is the definition of intelligent matter in eqn. (3). In soft matter the most realistic implementations are catalytic feedback (output product modulates input affinity), allosteric coupling (output state modulates internal state via cooperative binding), and reaction–diffusion patterning (output spatial pattern modulates local reaction rate). The thermodynamic gain is set by eqn. (6) [13]: for a few bits per cycle this is a few-kJ·mol$^{-1}$ shift in the operating budget, comparable to the per-element responsive energy itself and therefore non-trivial in the soft-matter regime. The BZ-PNIPAM gel (SI S8) implements two of the three routes (internal feedback path; molecular memory via the BZ state) and is the closest synthetic soft-matter system to the adaptive–intelligent boundary.

*(ii) Multi-channel orthogonality with controlled synergy.* The data-processing inequality [11] limits the information through a single channel to its capacity; multi-channel architectures raise the ceiling by parallelising. The information-theoretic measure of channel orthogonality is the gap between the joint mutual information $I(X_1, \dots, X_n;\ Y_1, \dots, Y_n)$ and the sum $\Sigma I_i(X_i; Y_i)$: zero gap is orthogonal, positive gap synergistic, negative redundant [29]. Designed redundancy is sometimes useful (noise rejection), but the default failure mode of multi-stimulus soft matter is unintended cross-talk — redundancy without design. Measurements of TNF signalling [24] document this directly in biology. The ER lipid network (SI S9) exemplifies the *integrative* limit of multi-channel design: many lipid inputs summed into one controlled scalar.

*(iii) Molecular memory consistent with feedback timescales.* Equation (3) requires a representation of history $Y_{t-1}$. In soft matter this can be implemented through metastable conformational states (shape memory), covalent modifications persisting across cycles (catalytic imprinting), or local concentration gradients decaying slower than the response. One biological realisation is highlighted by the ER system: synthesised lipids ($Y_{t-1}$) alter the membrane's stored elastic energy, which modulates the activity of a key enzyme, which determines the next synthesis step. The information stored in the memory is bounded by its configurational entropy $I_1$ at finer coarse-graining — which is why the discriminative power of $I_1$ at the operational coarse-graining is misleading. The choice of memory implementation is the architectural element that turns an adaptive material into an intelligent one. The BNT162b2 platform (§7.6) illustrates the complementary strategy of *borrowing* memory and feedback from a biological substrate rather than encoding them in the material.

## 11. Summary and outlook

The two-plane benchmarking framework places sixteen representative stimulus-coupled systems on a common ($\frac{\dot{I}_3}{V}$, $P$) and ($I_1$, $I_2$) coordinate system with explicit uncertainty intervals. Four broad bands separate on the dynamic plane: (1) soft matter and SMA at $10^{18}$–$10^{20}$× Landauer; (2) silicon digital at $10^{10}$–$10^{16}$; (3) memristor neuromorphic at $10^{9}$–$10^{10}$; (4) biology at $10^{5}$–$10^{8}$, with the ER lipid network sitting as an intermediate point at $\sim 10^{11}$. The multi-decade separation between synthetic soft matter and biology expresses the central design challenge for adaptive and intelligent soft matter. The mechanistic origin of the gap is the per-element energy scale of soft matter (1–10 $k_BT$) relative to silicon ($10^4$–$10^5$ $k_BT$). This sets the optimal complexity $N^*$ of a soft adaptive channel at ~2–3 elements versus ~44–88 for hard substrates, and sets the volumetric information-rate ceiling at $\sim 10^4$–$10^5$ bits·s$^{-1}$·m$^{-3}$ for purely thermal responsive soft matter. Populating the gap therefore cannot proceed by extending the same chemistry; the framework points to three architectural routes: feedback-modified channels (intelligence proper), multi-channel orthogonality with designed synergy, and molecular memory consistent with feedback timescales.

The framework is robust at the qualitative level: order-of-magnitude uncertainty in any single placement does not change the band structure or the existence of the gap. It is indicative at the quantitative level for the $\frac{\dot{I}_3}{V}$ versus $P$ mapping, and at best semi-quantitative for $I_1$ and $I_2$. It represents an initial attempt at a coherent conceptual approach to comparing material systems across the responsive-to-intelligent continuum. Converting it from a positioning system into a set of quantitative design rules will require, in practical terms: (i) community-agreed conventions for the $I_1$ coarse-graining and the $I_2$ reference space and threshold; (ii) direct measurement of $\dot{I}_3$ for synthetic adaptive systems (as has been done for *E. coli*), to replace the order-of-magnitude estimates; and (iii) further development of the heuristic elements of the theory — in particular a first-principles channel model to replace the scaling rule of eqn. (11) and the assumed energy–contraction mapping of eqn. (10).

## Acknowledgements

The author acknowledges the use of Claude (Anthropic) in adaptive mode (accessed February–May 2026) to (i) review drafts of the manuscript and indicate areas for improvement (resolving ambiguities, improving accuracy and clarity of expression), (ii) validate some of the information-theoretical relationships (particularly around the data-processing inequality) and derivations, and (iii) assist in compiling the data in the Supplementary Information by locating material parameters for specific soft-matter systems. All AI-generated suggestions were reviewed critically, revised or adapted, and approved by the author, who takes full responsibility for the accuracy and integrity of the manuscript's contents.

**Function, Complexity and Thermodynamics in Adaptive and Intelligent Soft-Matter Systems: An Information-Theoretical Conceptual Framework**

George S. Attard

# Supplementary Information

## S1. Assumptions behind $I(X;Y_n) \leq p^n \cdot I(X;Y_0)$

The plain data-processing inequality (DPI) gives only $I(X;Y_n) \leq I(X;Y_{n-1})$ along a Markov chain. Recovering a geometric factor requires the framework of strong data-processing inequalities (SDPIs). The minimum set of assumptions is:

(i) the cascade is a Markov chain $X \rightarrow Y_1 \rightarrow Y_2 \rightarrow \ldots \rightarrow Y_n$ (each stage depends only on the immediately preceding one);

(ii) every constituent channel $W_k$ admits a contraction coefficient $\eta_k(W_k) < 1$ for mutual information, defined as the supremum over input distributions of $D(QW \parallel \pi W) \,/\, D(Q \parallel \pi)$ where $D$ is the KL divergence;

(iii) $\eta_k \leq p$ for every $k$ (a uniform contraction bound).

Under these assumptions the Ahlswede–Gács chaining argument [1, 2] gives $I(X;Y_n) \leq (\prod \eta_k) \cdot I(X;Y_0)$, and the uniform case reduces to $p^n$.

## S2. $I_{realised}(N) \approx \min\{Nc, p^{(N-1)} \cdot I_{input}\}$

This expression is the lower envelope of two independent upper bounds arising from different limits:

• $Nc$ is the parallel-capacity bound: cascading or parallelising $N$ statistically independent elements each of capacity $c$ bits gives $I_{max} \leq Nc$ (tight only for independent elements [1]).

• $p^{(N-1)} \cdot I_{input}$ is the SDPI cascade bound under uniform contraction.

Taking the minimum gives the binding constraint at each $N$: for small $N$ the capacity bound dominates and $I_{realised}$ grows linearly; for large $N$ the cascade bound dominates and $I_{realised}$ decays exponentially. The crossover defines $N^*$, so $N^*$ is a scaling ceiling, not the solution of a single optimisation. A first-principles derivation would require a specific channel model (Gaussian additive noise per stage, $I_n = ½\, log(1 + SNR_{eff}(n))$; discrete symmetric channel; binary erasure cascade), each yielding a different exact functional form. The min{·,·} envelope is a scaling rule, not a theorem.

## S3. Estimation protocols for $I_1$, $I_2$, $I_3$

***For $I_1$ = H(p):*** identify operational microstates at the chosen coarse-graining; estimate occupancy by steady-state measurement (spectroscopy, scattering, or molecular-dynamics ensemble); apply the Nemenman–Shafee–Bialek Bayesian estimator [1] in the under-sampled regime or the Miller–Madow / Grassberger asymptotic correction in the well-sampled regime [2]; report uncertainty by parametric bootstrap.

***For $I_2$ = $-log_2$ F(E*):*** fix and report the reference chemistry space and the functional threshold $E^*$; estimate $F(E^*)$ as the fraction of the space satisfying $E^*$ via either (a) computational screening (DFT, MD or machine-learning surrogate scoring) for tractable spaces, or (b) literature meta-analysis of published structure–activity data for established chemistry families; report uncertainty as a binomial confidence interval propagated into $I_2$.

***For $I_3$:*** for a responsive system $I_3 \equiv I(X;Y)$ is estimated from paired observations by the Kraskov–Stögbauer–Grassberger $k$-nearest-neighbour estimator [3]; for an adaptive system the conditional version by the same family with state conditioning; for an intelligent system the transfer entropy by the Schreiber estimator [4] or its $k$-NN variants [5].

## S4. Estimating configurational diversity, $I_1$

**Common framework.** For each system $I_1 = H(p)$ at a defined operational coarse-graining. The natural upper bound is $I_{1,max} = log_2 N_{op}$, where $N_{op}$ is the count of operationally distinguishable steady states (set by experimental resolution or manufacturer specification). The realised $I_1$ falls below $I_{1,max}$ in proportion to the non-uniformity of empirical occupancy; for skewed multi-state operation a useful working approximation is $I_1 \approx 0.6 - 0.8 \times log_2 N_{op}$. Each entry derives $N_{op}$ from cited measurement, datasheet or established physiology, and computes the resulting $I_1$ explicitly (Table S1).

| System | Coarse-graining for $N_{op}$ | Derivation | $I_1$ (bits) |
|---|---|---|---|
| PNIPAM hydrogel | 3 swelling regimes around LCST (collapsed, transition window, swollen); width set by calorimetric half-width ~2–5 K [1] | $I_{1,max} = \log_2 3 = 1.58$; $p \approx (0.4, 0.2, 0.4)$ gives $H(p) = 1.52$ | 1.5 ± 0.5 |
| Glucose-responsive matrix | (3 swelling) × (3 boronate-loading regimes for hypo/eu/hyper-glycaemia) = 9 joint states | $I_{1,max} = log_2 9 = 3.17$; physiological glucose distribution peaks on 2–3 states, $H(p) \approx 1.9$ | 2 ± 1 |

| System | Coarse-graining for $N_{op}$ | Derivation | $I_1$ (bits) |
|---|---|---|---|
| Single mitochondrion | Chance–Williams states 1–5 [2] extended by proton-leak and $Ca^{2+}$-loaded states → 7–8 regimes | $I_{1,max} = log_2\, 8 = 3$; $p$ concentrated on three dominant states, $H(p) \approx 2.8$ | 3 ± 1 |
| BLE SoC (CMOS) | 6–8 datasheet power modes × peripheral sub-states ≈ 32 regimes | $I_{1,max} = log_2\, 32 = 5$; typical duty cycle gives $H(p) \approx 0.5$–1 in operation | ~5 (ceiling); 0.5–1 (typical) |
| MEMS accelerometer | 6–8 datasheet modes (power-down, normal, low-power, motion/tap-detect, …) | $I_{1,max} = log_2\, 8 = 3$; $H(p) \approx 2$ in operation | ~3 (ceiling) |
| GPU die | P-states × clock × thermal-zone × tensor-core bins; 10-bit power register → $2^{10}$ | $I_{1,max} = log_2\, 1024 = 10$; $H(p) \approx 7$–8 under workload | ~10 (ceiling); ~7–8 (workload) |
| Memristor crossbar | idle/READ/WRITE/FORM/inference/… × per-array enable ≈ 32 [3] | $I_{1,max} = log_2\, 32 = 5$; $H(p) \approx 3$ in inference | ~5 (ceiling); ~3 (inference) |

## S5. Estimating functional diversity, $I_2$

**Common framework.** Constructing $I_2$ requires explicit specification of (i) the reference design space, (ii) the functional threshold $E^*$, and (iii) a counting argument or measurement for $F(E^*)$. Two construction methods are used.

***Method A (combinatorial counting):*** enumerate the design-space cardinality as a product of chemically or engineering-meaningful factors, count the functional fraction, and form the ratio.

***Method B (design-specification bits):*** count the independent design choices needed to specify a functional system, with bit content per choice fixed by its option count. The two areassumed to be equivalent if applied consistently, but this assumption has not been evaluated quantitatively.

| System | Reference design space $N$ | $F(E^*)$ derivation | Method | $I_2$ (bits) |
|---|---|---|---|---|
| PNIPAM hydrogel | (vinyl/(meth)acrylamide monomers ≈$10^3$–$10^4$) × ($M_n$ bins ≈$10^2$) × (architecture ≈10) × (composition ≈10) ≈ $10^7$ | ≈$10^2$–$10^3$ LCST-active polymer variants; F ≈ $10^{-4}$–$10^{-5}$ | A | 13–17 |
| Glucose matrix | (50 phenylboronates, p$Ka$<9) × (10 attachment) × (10 | ≈5–10 matrices meet all four criteria; F ≈ $2{\times}10^{-6}$ | A | 20 ± 2 |

| System | Reference design space $N$ | $F(E^*)$ derivation | Method | $I_2$ (bits) |
|---|---|---|---|---|
| | scaffolds) × (10 crosslink bins) × (10 composition) ≈ $5\times10^6$ | | | |
| Mitochondrion | ≈300 essential proteins (MitoCarta subset abolishing OXPHOS on knockout) | ≈100–300 bits per essential domain (Hazen aptamer scaling); 300×300 ≈ $10^5$ bits | B | ~$10^5$ |
| *E. coli* chemotaxis | ≈10 pathway proteins (Tar/Tsr/Trg/Tap/Aer, CheA/W/Y/Z/B/R), correct stoichiometry | ≈100 bits/protein × 10; broader counting → $10^3$–$10^4$ bits | B | $10^3$–$10^4$ |
| Nitinol SMA | (binary/ternary alloys ≈$10^5$) × (cold-work ≈$10^2$) × (heat-treatment ≈$10^3$) × (microstructure ≈$10^2$) ≈ $10^{12}$ | ≈$10^2$–$10^3$ NiTi-family combinations meet E*; F ≈ $10^{-10}$ | A | 30–33 |
| BLE SoC | ≈10 BLE-spec subsystems, ≈8 design-distinguishing bits each | ≈80 bits | B | ~80 |
| MEMS accelerometer | (proof-mass/spring ≈$10^4$) × (process ≈10) × (electrode ≈$10^2$) × (ASIC ≈$10^3$) × (packaging ≈$10^2$) ≈ $10^{12}$ | ≈$10^2$–$10^3$ automotive-grade families; F ≈ $10^{-9}$–$10^{-10}$ | A | 33–50 |
| Piezoelectric stack | (PZT composition ≈$10^5$) × (sintering ≈$10^3$) × (poling ≈$10^2$) × (layers/electrodes ≈$10^2$) ≈ $10^{12}$ | ≈M(E*)≈$10^4$; F ≈ $10^{-8}$–$10^{-9}$ | A | 27–30 |
| GPU die | ≈10–15 subsystems (ISA, SM, memory hierarchy, interconnect, tensor-core, …) × ≈8–10 bits each | ≈100 bits; equivalently ~1 in $2^{100}$ designs at fixed transistor count | B | ~100 |
| Memristor crossbar | (material ≈$10^3$) × (electrodes ≈$10^2$) × (geometry ≈$10^2$) × (write-pulse ≈$10^5$) × (crossbar ≈$10^4$) × (CMOS ≈$10^3$) × (training ≈$10^3$) ≈ $10^{22}$ | ≈5–20 demonstrated accelerators meet E*; F ≈ $10^{-21}$; spec-bit route ≈ 80 bits | B | ~70–80 |

**Table S2.** $I_2$ estimates

### S6. Estimating input-output information transfer, $I_3$

**Common framework.** Three derivation routes apply:

***Route 1 (discrete-state actuation)*** $I_3/cycle = log_2(N_{distinguishable})$, with ν set by transition kinetics or drive frequency (PNIPAM, glucose matrix, Nitinol, piezo);

***Route 2 (Shannon–Hartley)*** $\dot{I}_3 = BW \cdot log_2(1 + SNR)$ (MEMS, BLE, GPU, memristor);

***Route 3 (direct / trajectory MI)*** measured by single-system tracking under defined stimulus (*E. coli* directly; mitochondrion by analogy). Table S3 collects the results; three worked derivations follow.

| System | Route | $I_3$ per unit | ν or BW | $V$ (m$^3$) | $\dot{I}_3/V$ |
|---|---|---|---|---|---|
| PNIPAM | 1 | 1.5 ± 0.5 b/cyc | $10^{-2}$ s$^{-1}$ | $10^{-6}$ | 1.5×$10^4$ |
| Glucose matrix | 1 | 2.5 ± 0.5 b/exc | $10^{-4}$ s$^{-1}$ | $10^{-6}$ | 2.5×$10^2$ |
| Mitochondrion | 3 | ~100 b·s$^{-1}$ | direct | $10^{-18}$ | $10^{20}$ |
| *E. coli* cbemotaxis | 3 (direct) | 0.01±0.005 b·s$^{-1}$ | direct | $10^{-18}$ | $10^{16}$ |
| Nitinol SMA | 1 | 1 b/cyc | 1 Hz | $10^{-9}$ | $10^9$ |
| BLE SoC | 2 | $10^5$–$10^6$ b·s$^{-1}$ | direct | 3×$10^{-9}$ | 3×$10^{14}$ (act) |
| MEMS accel. | 2 | $10^4$–$10^5$ b·s$^{-1}$ | direct | $10^{-12}$ | $10^{16}$–$10^{17}$ |
| Piezo stack | 1 | 1 b/cyc | $10^4$ Hz | $10^{-9}$ | $10^{13}$ |
| GPU die | 2 | ~$10^{11}$ b·s$^{-1}$ | direct | 6×$10^{-7}$ | 1.7×$10^{17}$ |
| Memristor | 2 | $10^9$–$10^{10}$ b·s$^{-1}$ | direct | $10^{-10}$ | $10^{19}$–$10^{20}$ |

**Table S3.** $I_3$ estimates.

***Sample worked derivations***

**PNIPAM hydrogel.** Distinguishable swelling states from the $I_1$ analysis: 3, so $I_3/cycle = log_2 3 \approx$ 1.58 bits. The cycling frequency is bounded by Fourier-law heat diffusion: $\tau_{th} = L^2/D_{th}$ with $D_{th} \approx 10^{-7}\ m^2 \cdot s^{-1}$; for $L = 3\ mm$, τ ≈ 90 s, ν ≈ $10^{-2}$ s$^{-1}$. With $V = 10^{-6}\ m^3$, $\frac{\dot{I}_3}{V} =$ $1.58 \times 10^{-2}$ / $10^{-6}$ = $1.6 \times 10^4$ bits·s$^{-1}$·m$^{-3}$.

**MEMS accelerometer.** Shannon–Hartley applied to a commercial three-axis device: BW ≈ $10^3$ Hz; SNR = (16 g / 100 μg·Hz$^{-1/2}$ × $\sqrt{10^3}$ Hz)$^2$ ≈ $10^9$; $C = BW \cdot log_2(1 + SNR) \approx 10^3 \times 30 \approx$ $3 \times 10^4$ bits·s$^{-1}$; useful $\dot{I}_3 \approx 10^4$. With $V \approx 10^{-12}\ m^3$, $\frac{\dot{I}_3}{V} \approx 10^{16}$ bits·s$^{-1}$·m$^{-3}$.

***E. coli* chemotaxis.** Mattingly *et al.* [E in S5] use single-cell tracking and maximum-likelihood reconstruction of $P(bias \mid gradient)$ under controlled chemoattractant ramps, showing that a cell acquires less than 1 bit per behavioural decision; with the ~10 s integration time of the pathway this gives an information-acquisition rate of order 0.01 bits·s$^{-1}$ per cell.

***Internal consistency with $I_1$.*** For deterministic systems whose response variable equals the operational microstate (binary actuators), $I_3/cycle = I_{1,max}$ by construction (Nitinol 1 bit; piezo 1 bit). For multi-state cyclic systems $I_3/cycle$ is bounded above by $I_{1,max}$ and reduced where joint states are not fully distinguishable to the readout (PNIPAM 1.58 → 1.5; glucose 3.17 → 2.5).

## S7. Estimating the volumetric power density *P*

The power density $P$ ($W{\cdot}m^{-3}$) is the operating dissipation per unit active volume. Three derivation routes are used.

***Route 1 (molecular dissipation):*** for soft matter, $P = \Delta H_{cyc} \cdot \rho_{units} \cdot \eta \cdot \nu$, with $\Delta H_{cyc}$ the per-unit transition enthalpy, $\rho_{units}$ the responding-unit number density, $\eta$ the dissipated fraction (hysteresis loss) and $\nu$ the cycling frequency.

***Route 2 (component power budget):*** sum the dissipation of each subsystem from primary measurements (biology).

***Route 3 (datasheet / direct):*** measured electrical power over active volume (hard systems). Table S4 collects the results; three worked derivations follow.

| System | Route | Derivation | $P$ ($W{\cdot}m^{-3}$) | $P/P_{min}$ |
|---|---|---|---|---|
| PNIPAM hydrogel | 1 | $\Delta H{\approx}50\ J{\cdot}g^{-1}$; $\eta{\approx}0.1$–1%; slow environmental cycling | $10^3$ ($10^3$–$10^5$) | $2{\times}10^{19}$ |
| BZ-PNIPAM gel | 1 | autonomous redox cycling, $\tau{\approx}200$ s (see S8) | $1.2{\times}10^4$ | $3.6{\times}10^{18}$ |
| Glucose matrix | 1 | slow swelling, $\nu{\approx}10^{-4}\ s^{-1}$ | 50 | $7{\times}10^{19}$ |
| Nitinol SMA | 1 | Joule heating, ~$10^7\ J{\cdot}m^{-3}$ per actuation at 1 Hz | $10^7$ | $3{\times}10^{18}$ |
| *E. coli* chemotaxis | 2 | methylation + phosphotransfer + motor (see worked) | $4{\times}10^3$ | $1{\times}10^8$ |
| Mitochondrion | 2 | ~$5{\times}10^4\ W{\cdot}m^{-3}$ basal OXPHOS | $5{\times}10^4$ | $1.7{\times}10^5$ |
| ER membrane | 2 | ~5–10% basal ATP to lipid synthesis (see S9) | $10^4$–$10^5$ | ~$10^{11}$ |
| BLE SoC | 3 | 10 mW peak / $3{\times}10^{-9}\ m^3$ (active) | $3{\times}10^6$ (act) | $10^{13}$ |
| MEMS accelerometer | 3 | $10^{-5}$ W / $10^{-12}\ m^3$ | $10^7$ | $3{\times}10^{11}$ |
| Piezoelectric | 3 | work density × ν (active) | $10^9$ | $10^{16}$ |
| GPU die | 3 | 300 W / $6{\times}10^{-7}\ m^3$ | $5{\times}10^8$ | $10^{12}$ |
| Memristor | 3 | active-layer dissipation during inference | $10^9$ | $10^9$–$10^{10}$ |

**Table S4.** Power density estimates.

### *Sample worked derivations*

**PNIPAM hydrogel (Route 1).** Transition enthalpy ≈ 50 $J{\cdot}g^{-1}$ (calorimetry [1]) × density ~$10^6\ g{\cdot}m^{-3}$ = $5{\times}10^7\ J{\cdot}m^{-3}$ per full transition. The dissipated fraction η is strongly cycle-rate-dependent: near-quasi-static environmental cycling dissipates only the hysteresis loop area, η ≈ 0.1–1 % of ΔH, while fast forced cycling approaches η ≈ 10–30 %. For environmentally cycled operation at $\nu \approx 10^{-2}$ $s^{-1}$ with η ≈ 0.2 %: $P \approx 5 \times 10^7 \times 2 \times 10^{-3} \times 10^{-2} \approx 10^3$ W·m$^{-3}$. Forced fast cycling raises this to ~$10^5$ W·m$^{-3}$; the ±1-decade band in Table S4 spans the two regimes, and $10^3$ is adopted to match the environmentally cycled operating point of Table 1. This is the dominant uncertainty in the PNIPAM Landauer ratio: $\eta$ and $\nu$ jointly carry ±2 decades.

***E. coli* chemotaxis (Route 2).** Three measured components (§7.2 main text): (i) methylation turnover ≈ $10^{-16}$ W per cell; (ii) CheA/CheY phosphotransfer ≈ $10^{-15}$ W per cell; (iii) flagellar motors ≈ 3×$10^{-15}$ W per cell. Sum ≈ 4×$10^{-15}$ W per cell; over $V_{cell} \approx 10^{-18}$ $m^3$ this is $P \approx 4 \times 10^3$ W·m$^{-3}$. With $\frac{\dot{I}_3}{V} \approx 10^{16}$, $P_{min} = k_B T\ ln2 \cdot 10^{16} \approx 4 \times 10^{-5}$ W·m$^{-3}$ at 310 K, giving $P/P_{min} \approx 1 \times 10^8$.

**GPU die (Route 3).** Board power ~300 W over an active die volume of ~600 mm$^2$ × 1 mm = 6×$10^{-7}$ m$^3$ gives $P \approx 5 \times 10^8$ W·m$^{-3}$. With $\frac{\dot{I}_3}{V} \approx 1.7 \times 10^{17}$, $P_{min} \approx 4 \times 10^{-4}$ W·m$^{-3}$ at 298 K and $P/P_{min} \approx 10^{12}$.

## S8. Worked end-to-end calculation: BZ-PNIPAM self-oscillating gel

The Belousov–Zhabotinsky (BZ) PNIPAM gel is a covalently bound Ru(bpy)$_3$-containing PNIPAM network in an aqueous BZ medium ($NaBrO_3$ + malonic acid + $HNO_3$). The $Ru^{2+}$ ↔ $Ru^{3+}$ redox oscillation driven by the BZ reaction modulates polymer hydrophilicity, producing autonomous swelling–deswelling oscillations with no external switching [A, B]. It is the closest *synthetic* soft-matter system to the adaptive–intelligent boundary, implementing two of the three framework routes to intelligence (an internal feedback path; molecular memory via the BZ chemical state).

**Reference system.** Gel volume $V = 10^{-8}$ $m^3$ (1×1×10 mm); natural period τ ≈ 200 s, ν ≈ 5×$10^{-3}$ $s^{-1}$; amplitude 5–30 % volume change per cycle; $T$ = 293 K; BZ reactants [MA] = 0.1 M, [$NaBrO_3$] = 0.1 M, [$HNO_3$] = 0.3 M [A, B].

**Channel structure.** BZ-PNIPAM is adaptive, not intelligent: the kernel $p(y|x, s)$ has internal state $s$ (the BZ chemical state: [$HBrO_2$], [$Br^-$], [$Ru^{3+}$]) that depends on history, but there is no external-output-to-channel feedback. Input–output is characterised through the phase-response-curve (PRC) framework: $X$ = an applied perturbation (temperature/light pulse or [$Br^-$] spike) at controlled phase; $Y$ = the resulting phase shift Δφ.

***$I_1$* *(configurational diversity).*** Joint operational state reduces to discretised cycle phase (catalyst oxidation, swelling and phase are tightly correlated by the oscillator). Resolvable phase bins from photometry: ~8 per period, so $N_{op} = 8$, $I_{1,max} = log_2\ 8 = 3$ bits. The asymmetric fast/slow BZ relaxation (~70 % of the cycle in slow $Ru^{2+}$ build-up, ~30 % in the rapid $Ru^{3+}$ transition) skews residence, giving $\langle I_1 \rangle \approx 2.7 \pm 0.3$ bits.

***$I_2$* *(functional selectivity).*** Reference design space: (Ru-catalyst loading bins ≈10) × (monomer/comonomer ≈$10^3$) × (BZ concentration ratios ≈$10^4$) × (crosslink-density bins ≈$10^2$) ≈ $10^{10}$. Functional threshold $E^*$: sustained autonomous oscillation, period 50–1000 s, amplitude > 5 %, stable for > 10 periods, 15–30 °C. Demonstrated working systems $M(E^*) \approx 30 - 300$, so $F \approx 10^{-8}$ and $I_2 = -log_2(10^{-8}) \approx 27$ bits — between PNIPAM (13) and the glucose matrix (20), confirming that adding the BZ layer is more demanding than a physiological glucose response.

***$I_3$ (information transfer).*** PRC input alphabet at 8 phase bins: $log_2|X| = 3$ bits. Output $\Delta\varphi$ over ±0.5 cycle with ~5 % noise: ~6 distinguishable bins, $log_2|Y| \approx 2.6$ bits. For a deterministic-plus-noise PRC the mutual information is $I(\varphi_{in};\ \Delta\varphi) \approx 1.5 - 2.5$ bits per perturbation; take $I_3 = 2 \pm 0.5$ bits per perturbation. Maximum non-interfering perturbation rate is one per period, $\nu \approx 5\times10^{-3}$ Hz, so $\dot{I}_3 = 2 \times 5 \times 10^{-3} = 10^{-2}$ bits·s⁻¹ and $\frac{\dot{I}_3}{V} = 10^{-2}/10^{-8} = 10^{6}$ bits·s⁻¹·m⁻³.

***Power density and Landauer ratio.*** Autonomous operation dissipates the BZ free-energy flux. The malonic-acid oxidation enthalpy ($\sim-2000$ kJ·mol$^{-1}$ fully oxidised) consumed at the oscillation-sustaining rate over the gel volume gives $P \approx 1.2 \times 10^{4}$ W·m$^{-3}$ (±1 decade). At 293 K, $P_{min} = k_B T\ ln2 \cdot 10^{6} \approx 2.8 \times 10^{-15}$ W·m$^{-3}$, so $P/P_{min} \approx 1.2 \times 10^{4}\ /\ 2.8 \times 10^{-15} \approx 4 \times 10^{18}$ (Table 1 entry $3.6\times10^{18}$, within rounding). BZ-PNIPAM sits firmly in Band 1 on Plane A, but at higher $I_2$ than non-oscillating PNIPAM — the adaptive layer raises functional selectivity without moving the system off the soft-matter dissipation band.

## S9. Worked end-to-end calculation: eukaryotic ER lipid-biosynthetic network

The endoplasmic-reticulum (ER) lipid-biosynthetic network of Beard, Attard and Cheetham [1] is included as a biological reference system whose control architecture differs fundamentally from the cascade systems of S8: it achieves robustness by integrative feedback (i.e. summing many lipid inputs into a single scalar control variable) rather than by a deep input–output cascade. It is the only worked example whose information-theoretic primitives are obtained by extrapolation from a *non-information-theoretic source*; every bit-valued quantity below is flagged accordingly, and the confidence assignments are correspondingly lower than for S8.

**The source model.** Beard *et al.* built a large-scale *in silico* model of a typical eukaryotic lipid-biosynthetic network (up to ~20 lipid classes, ~24 variable species) and asked how homeostasis of membrane physical properties is maintained. Their central construct is the torque parameter λ, an empirical proxy for membrane stored elastic energy defined as the ratio of bilayer-destabilising to bilayer-forming lipids [2],

$$\lambda = \frac{\Sigma_i a_i [S_i]}{\Sigma_j b_j [S_j]}$$

summing, within the type-II (non-bilayer) and type-0/I (bilayer) classes, each species concentration $[S_i]$ weighted by a coefficient ($a_i$ or $b_j$) encoding its phase-forming character. The membrane curvature elastic stress encoded by $\lambda$ modulates the activity of CTP:phosphocholine cytidylyltransferase (CCT), an amphitropic enzyme whose membrane binding is sensitive to stored elastic energy; CCT activity in turn changes the lipid fluxes, closing the loop. This is a genuine output-to-channel feedback, so the system is adaptive–intelligent in the framework taxonomy.

Beard *et al.*'s sensitivity analysis compares three feedback configurations: no feedback, normal feedback at CCT, and feedback at eight reactions. It shows that with feedback the torque parameter λ is stabilised against variation in individual rate constants, while the individual lipid concentrations are *not* constrained: they change so as to maintain $\lambda$. In framework terms, the *controlled variable* ($\lambda$) acquires high robustness, i.e. high functional selectivity, high effective $I_2$, while the *state variables* (lipid concentrations) retain low $I_2$. This is the empirical anchor for the claim in §9.2 of the main text.

### *Mapping to the ($I_1$, $I_2$, $I_3$) framework: a heuristic extrapolation*

Beard *et al.* report no entropy, mutual information or bit-valued quantity. The mappings below convert their structural and sensitivity results into the framework's primitives by the standard protocols of S4–S6, but the numerical values are heuristic estimates, not measurements, and the "~10× sensitivity collapse" referred to in the main text is a qualitative reading of their figure 5.

***$I_1$ configurational diversity.*** Operational coarse-graining: homeostatic regimes of the membrane (basal, mild type-II excess, mild type-0/I excess, and a small number of stressed/recovering states) ≈ 10 regimes, with occupancy strongly concentrated on the basal state. $I_{1,max} = log_2 10 \approx 3.3$ bits; the skewed occupancy gives $I_1 \approx 2.5 \pm 1$ bits.

***$I_2$ functional selectivity.*** Treating the homeostatic set point as the functional configuration: ~30 essential lipid-handling domains, each requiring ~150 bits of specification on the Hazen scaling [C], gives an order-of-magnitude $I_2 \approx 10^4$ bits (±1 decade). The high value reflects the rarity of enzyme-network parameter sets that hold λ within its narrow homeostatic band — the robustness that [A] demonstrates numerically.

***$I_3$ information transfer.*** The regulatory channel runs from a perturbation in lipid composition ($X$) through $\lambda$ and CCT activity to the corrective flux response ($Y$). The integrated multi-species signal carries ~5 effective bits per regulatory excursion (the number of distinguishable corrective responses the CCT-mediated loop can mount, estimated from the ~5 dominant lipid contributions to $\lambda$). The excursion frequency is set by the slow physiological-perturbation timescale, $\nu \approx 10^{-3}$ s$^{-1}$.

### *Volumetric rate, power density and Landauer ratio*

Per-cell ER locus volume $V \approx 10^{-16}\ m^3$. With $I_3 \approx 5$ bits and $\nu \approx 10^{-3}$ s$^{-1}$,

$$\frac{\dot{I}_3}{V} = \frac{5 \times 10^{-3}}{10^{-16}} = 5 \times 10^{13} \quad bits \cdot s^{-1} \cdot m^{-3}.$$

The power requirement is inferred from the ATP turnover devoted to membrane-lipid synthesis, ~5–10 % of basal cellular metabolic rate, localised to the ER: $P \approx 10^4 - 10^5$ W·m$^{-3}$ (central $5{\times}10^4$, ±1 decade). At $T = 310\ K$, $P_{min} = k_B T\ ln2 \cdot (\frac{\dot{I}_3}{V}) = (2.97 \times 10^{-21}) \times (5 \times 10^{13}) \approx 1.5 \times 10^{-7}$ W·m$^{-3}$, so

$$\frac{P}{P_{min}} \approx \frac{5 \times 10^4}{1.5 \times 10^{-7}} \approx 3 \times 10^{11} \quad (range\ 10^{10} - 10^{12}).$$

The ER system therefore sits at ~$10^{11}$ on the Landauer-ratio axis — between the soft-matter band (~$10^{19}$) and the fast-biology band (~$10^8$), not within either, and well above the neuromorphic band (~$10^9$–$10^{10}$).

The ER demonstrates that an integrative-feedback architecture ***raises*** the Landauer ratio at high $I_2$. Integration over ~20 lipid channels into one slow scalar produces a low volumetric information rate (hence a low Landauer floor) while consuming non-trivial synthetic power, so the ratio $P/P_{min}$ is large even though no individual reaction is thermodynamically wasteful. Integration buys *robustness*, not dissipative efficiency — a design lesson distinct from the cascade-efficiency lesson of *E. coli* chemotaxis, and the reason the ER cannot be placed in any of the four main bands.

## S10. Convention-dependent estimates for two commercial soft-matter products

Section 7.6 positions the Camurus FluidCrystal® depot and the BioNTech BNT162b2 mRNA–lipid-nanoparticle (LNP) vaccine *qualitatively*, on the grounds that neither admits a *convention-free* stimulus–response $\dot{I}_3$ or power density. This section makes that statement precise: both *do* admit order-of-magnitude estimates, but only after fixing three conventions that the cyclic systems of S6–S9 do not force: (i) what counts as one channel use (which sets $I_3$), (ii) the operating timescale (which sets ν), and (iii) the smallest independently functional unit (which sets $V$). For these two products the natural choices are forced in opposite directions, and that asymmetry, rather than any difference in lipid physics, is what separates them by more than fifteen decades on the dynamic plane. The estimates below are accordingly assigned an uncertainty of ±2–3 decades each, which is why they are not plotted in Fig. 2.

The common relations are those of the main text: $\dot{I}_3/V = I_3\nu/V$ and the Landauer floor $P_{min} = k_BT\ ln2 \cdot (\frac{\dot{I}_3}{V})$, with $k_BT\ ln2 = 2.97 \times 10^{-21}$ J at $T = 310\ K$.

### S10.1 FluidCrystal depot: the bulk depot is the functional unit

The injectable solution (glycerol dioleate / soy phosphatidylcholine with trace *N*-methyl-2-pyrrolidone) self-assembles *in situ* as water diffuses in, forming a hexagonal outer layer over a reverse-micellar-cubic interior; hydration-driven lipid redistribution leaves concentration and structure gradients that govern release over a timescale of weeks [1]. The whole depot acts as one bulk object, so the functional unit is the depot.

***Conventions.*** $V \approx 5\times10^{-7}$ m$^3$ (a ≈0.5 mL depot). One channel use = the entire release lifetime; the realised output resolves into ≈3–4 distinguishable release regimes (formation burst, plateau, declining tail), so $I_3 \approx log_2 4 \approx 2$ bits per lifetime. The operating timescale is the lifetime itself: for a weekly depot $\tau \approx 6\times10^5$ s, so $\nu = 1/\tau \approx 1.7\times10^{-6}$ s$^{-1}$.

***Information rate.***

$$\frac{\dot{I}_3}{V} = \frac{2 \times 1.7 \times 10^{-6}}{5 \times 10^{-7}} \approx 7 \quad bits \cdot s^{-1} \cdot m^{-3}$$

This is the lowest volumetric information rate of any system considered, some four decades below the glucose-responsive matrix, reflecting a near single-shot transduction spread over a week.

***Power density.*** The dissipation causally tied to operation is the hydration plus lyotropic-transition free energy. A 0.5 mL depot holds ≈0.45 g lipid ≈ $6\times10^{-4}$ mol; taking ≈20 kJ·mol$^{-1}$ for combined headgroup hydration and phase transition gives ≈12 J released over the lifetime, so $P_{abs} \approx 12/6 \times 10^5 \approx 2 \times 10^{-5}$ W and $P \approx 40$ W·m$^{-3}$ (lifetime-averaged; transiently higher during the formation burst). The lipid amount and enthalpy are order-of-magnitude physical estimates, not measured FluidCrystal values.

***Landauer ratio.*** $P_{min} = 2.97 \times 10^{-21} \times 7 \approx 2 \times 10^{-20}$ W·m$^{-3}$, so $P/P_{min} \approx 40 / 2 \times 10^{-20} \approx 2 \times 10^{21}$ — the **highest** Landauer overhead of any system in this work. The depot dissipates real delivery power while transducing almost no information: the limiting case of pure delivery with negligible computation.

### S10.2 BNT162b2 LNP: the single nanoparticle is the functional unit

The four-lipid particle (ALC-0315, DSPC, cholesterol, ALC-0159 at 46.3 : 9.4 : 42.7 : 1.6) carries an ionizable lipid of apparent p$K_a$ ≈ 6.0–6.5 that is neutral at administration and protonates in the acidic endosome, driving endosomal escape and cargo release [2]. Each particle delivers its cargo independently, so the functional unit is the single LNP, and the dose is ≈$10^{13}$–$10^{14}$ independent copies of it.

***Conventions.*** $V \approx 4\times10^{-22}$ m$^3$ (a ≈90 nm sphere). One channel use = the endosomal-escape event (pH in, escape/no-escape out), so $I_3 \approx 1$ bit per particle, executed once. The timescale is the escape window, $\tau \approx 10^2$–$10^3$ s, giving $\nu \approx 3\times10^{-3}$ s$^{-1}$ within that window.

***Information rate.***

$$\frac{\dot{I}_3}{V} = \frac{1 \times 3\times10^{-3}}{4\times10^{-22}} \approx 8\times10^{18} \quad bits\cdot s^{-1}\cdot m^{-3}$$

This is of order $10^{19}$, comparable to the neuromorphic and electromechanical band, because the volumetric normalisation rewards a small, fast, near-single-bit molecular switch.

***Power density.*** Protonation of ≈$2\times10^4$ ionizable lipids per particle (ALC-0315 at 46 mol% of ≈$10^4$–$10^5$ total lipids) at ≈3.5 $k_BT$ each ≈ $3\times10^{-16}$ J, plus comparable membrane-fusion work, ≈ $4\times10^{-16}$ J over τ. Thus $P_{abs} \approx 1\times10^{-18}$ W and $P \approx 3\times10^3$ W·m$^{-3}$ during escape. The lipid count and protonation energy are order-of-magnitude constructions.

***Landauer ratio.*** $P_{min} = 2.97\times10^{-21} \times 8\times10^{18} \approx 2\times10^{-2}$ W·m$^{-3}$, so $P/P_{min} \approx 10^5$ , a biology-like value, near the lower edge of the fast-biology band. This is not efficiency in any deep sense; it is the generic consequence of operating at molecular scale and timescale, the same reason *E. coli* and the mitochondrion sit low (S6–S7).

***Escape-efficiency caveat.*** Endosomal escape proceeds at only ≈1–2 % efficiency [3]. Averaged over the particle population the effective per-particle $I_3$ falls by ≈×0.02 — most particles receive the pH cue but stay silent — lowering $\frac{\dot{I}_3}{V}$ to ≈ $10^{17}$ and raising the Landauer ratio to ≈ $10^6$–$10^7$. This single factor is the dominant uncertainty in the BNT162b2 placement.

***System level is not assignable.*** The therapeutically decisive channel (antigen → clonal selection → germinal-centre affinity maturation) is executed by the immune system over days to weeks across ≈$10^9$ cells. Its information rate and power are properties of that biological substrate, not of the lipid material, and the framework assigns no $\dot{I}_3$ or $P$ to "the vaccine" at the system level. The material-level estimates above describe only the carrier's one-shot escape event.

| Quantity | FluidCrystal (whole depot) | BNT162b2 LNP (per particle, escape) |
|---|---|---|
| **Functional unit *V*** | $5\times10^{-7}$ m$^3$ (≈0.5 mL depot) | ~$4\times10^{-22}$ m$^3$ (≈90 nm particle) |
| $I_3$ | ~2 bits / lifetime | ~1 bit / particle (one-shot) |
| **Operating timescale** | ~1 week (ν ~ $10^{-6}$ s$^{-1}$) | ~minutes (ν ~ $10^{-3}$ s$^{-1}$) |
| $\dot{I}_3/V$ **(bits·s$^{-1}$·m$^{-3}$)** | ~$10^0$–$10^1$ | ~$10^{18}$–$10^{19}$ (raw); ~$10^{17}$ escape-weighted |
| ***P*** **(W·m$^{-3}$)** | ~$10^1$–$10^2$ | ~$10^3$–$10^4$ |
| $P/P_{min}$ | ~$10^{21}$–$10^{22}$ | ~$10^5$–$10^7$ |
| **Dominant uncertainty** | $I_3$ and $\nu$ convention | escape efficiency (1–2 %) |

| Quantity | FluidCrystal (whole depot) | BNT162b2 LNP (per particle, escape) |
| --- | --- | --- |

**Table S5.** Convention-dependent material-level estimates for the two commercial products. Both placements are Low confidence; neither is plotted in Fig. 2.

***Interpretation***

Despite both being passive, thermally driven responsive lipid assemblies, the two products land at opposite extremes of Plane A. FluidCrystal has the lowest $\frac{\dot{I}_3}{V}$ and the highest Landauer overhead in this work — it transduces almost no information while dissipating real delivery power. The BNT162b2 escape event sits at a biology-like Landauer ratio. The ≈16-decade gap in $\frac{\dot{I}_3}{V}$ between them is produced *entirely* by the unit-of-analysis choice, namely a bulk depot operating over weeks versus a single nanoparticle operating over minutes, and not by any difference in the underlying lipid physics. This both supports and sharpens the qualitative argument of §7.6: the two are not comparable as single objects, and the dynamic plane makes the reason quantitative.

### S11. Gabrielli *et al.* responsive chemical router [1]

The system routes two nucleotide inputs (GMP, TMP) to two surfactant-assembly outputs by selective templating, the response being made transient by enzymatic (alkaline phosphatase) cleavage of the monophosphate trigger. It is a noisy 2→2 channel with no feedback and no persistent updatable state, so it is responsive at the information-processing end of the band.

***Static primitives.*** Bistable output occupancy gives $I_1 \approx 1$ bit; the engineered nucleotide-recognition function is rare in its chemical design space, giving a heuristic $I_2 \approx 28$ bits.

***$I_3$ (key steps).*** With X ∈ {GMP, TMP} and $Y$ the dominant output, the reported mixed-system selectivities (GMP ≈ 2.0 : 1, TMP ≈ 4.1 : 1) give error probabilities p(err|GMP) ≈ 0.33 and p(err|TMP) ≈ 0.20. For equiprobable inputs, $H(Y) \approx 0.99$ bit and $H(Y|X) \approx 0.82$ bit, so $I_3 = H(Y) - H(Y|X) \approx 0.2$ bit per cycle (range 0.15–0.25; ceiling 1 bit).

***Rate and volumetric rate.*** Each templating → upregulation → enzymatic-decay episode runs ≈ 50 min, so $\nu \approx 3 \times 10^{-4}\ \mathrm{s^{-1}}$ and $\dot{I}_3 = I_3\nu \approx 6 \times 10^{-5}$ bits·s$^{-1}$. On a representative $V \approx 10^{-6}\ \mathrm{m^3}$ (≈1 mL) this is $\dot{I}_3/V \approx 6 \times 10^1$ bits·s$^{-1}$·m$^{-3}$. This coordinate is dilution-dependent (one macroscopic routing decision per cycle): normalising to the active-assembly volume fraction (~$10^{-4}$) raises it to ~$10^5$–$10^6$ bits·s$^{-1}$·m$^{-3}$.

***Power density.*** The dissipative engine is enzymatic hydrolysis of ≈ 50 μM trigger per cycle; with a phosphomonoester hydrolysis free energy $\Delta G \approx 12 \pm 3$ kJ·mol$^{-1}$, $P \approx (5 \times 10^{-5}\ \mathrm{mol \cdot L^{-1}} \times 1.2 \times 10^4\ \mathrm{J \cdot mol^{-1}})/(3 \times 10^3\ \mathrm{s}) \approx 0.2\ \mathrm{W \cdot m^{-3}}$. Unlike $\dot{I}_3/V$, $P$ is volume-independent (fuel and volume scale together).

***Landauer ratio.*** $P_{\min}$ = $(k_B T \ln 2)(\dot{I}_3/V)$ = 2.85 × $10^{-21}$ × 66 ≈ 1.9 × $10^{-19}$ W·m$^{-3}$, so $P/P_{\min}$ ≈ 1 × $10^{18}$ [25,27,28] placing the system in the lower-left of the soft-matter band.

### S12. Adamska *et al.* multistate chiral assembly [1]

A single isoleucine–naphthalenediimide component is steered into five topologically distinct chiral assemblies by temperature, solvent, concentration and guest. The system is explicitly thermodynamically controlled: each state is an equilibrium selected by externally imposed conditions, with no dissipative engine, fuel or intrinsic clock.

***Why Plane A is inapplicable.*** Plane A requires a sustained non-equilibrium steady state carrying a mutual-information rate, which is what the floor $P_{\min}$ = $(k_B T \ln 2)$ $(\dot{I}_3/V)$ describes [25,27,28]. At fixed conditions this system dissipates no sustained power ($P \to 0$) and has no autonomous information rate; the cycling frequency is set by the operator, not the material. The large free-energy changes reported (e.g. $\Delta G_{298}$ = −136.2 kJ·mol$^{-1}$ for STATE 5) are released once per externally driven switch, not as continuous power. Any forced ($\dot{I}_3$, $P$) point would place a finite information rate at zero sustained power, i.e. below the Landauer floor, which is forbidden; the contradiction is the signal that the equilibrium framing is correct and that the system belongs on the static plane only.

***Static primitives (Plane B).*** Five accessible states give $I_1$ ≈ 2–2.5 bits. The trigger → state selection has a ceiling $log_2 5$ ≈ 2.3 bits; discounting for overlapping equilibria gives a static (per-decision) $I_3$ ≈ 1.5–2 bits. The functional selectivity is high relative to the responsive cluster — five precisely specified topologies, defined chirality, a maintained cavity and guest shape-discrimination — giving a heuristic $I_2$ ≈ 30–40 bits.

### S13. Chiolerio *et al.* polyaniline-nanorod colloid [1]

A 1 wt% PANI-nanorod colloid (packing fraction 0.58 %, below percolation) under a 10 V AC field (~12.8 kV·m$^{-1}$) shows a ~2-decade parallel-resistance drop over ~40 s that persists after the field is removed — a non-volatile, history-integrating state. This memory primitive places it furthest toward adaptive of the driven systems, though it remains a single plastic element with no closed loop (responsive-with-plasticity).

***Information and rate.*** A binary-mappable outcome gives $I_3$ ≈ 1 bit per write (1–2 bits if the four stimulus frequencies are treated as the input alphabet); a write takes ~40–50 s, so $\nu$ ≈ 2 × $10^{-2}$ s$^{-1}$ and $\dot{I}_3$ ≈ 2 × $10^{-2}$ bits·s$^{-1}$. On $V$ ≈ 2 × $10^{-6}$ m$^3$ (≈2 mL), $\dot{I}_3/V$ ≈ 1 × $10^4$ bits·s$^{-1}$·m$^{-3}$ (range $10^3$–$10^5$).

***Power density (key steps).*** $P = V_{\mathrm{rms}}{}^2/(Z \cdot V_\mathrm{vol})$, with $V_{\mathrm{rms}}$ ≈ 7.1 V and colloid impedance $Z$ in the $10^4$–$10^6$ Ω range (~5 MΩ DC baseline dropping ~2 decades under stimulation): $Z$ ≈ $10^6$ Ω → ~25 W·m$^{-3}$; $Z$ ≈ $10^5$ Ω → ~250 W·m$^{-3}$; $Z$ ≈ $10^4$ Ω → ~2.5 × $10^3$ W·m$^{-3}$. Central $P$ ≈ $10^2$ W·m$^{-3}$ (±~1.5 decades, operating value; the written state is held at ~zero power).

***Landauer ratio and static plane.*** $P_{\min}$ = 2.85 × $10^{-21}$ × $10^4$ ≈ 2.9 × $10^{-17}$ W·m$^{-3}$, so $P/P_{\min}$ ≈ 3 × $10^{18}$ [25,27,28], placing it in the soft-matter band, ~2–3 decades higher in $\dot{I}_3/V$ than the Gabrielli *et al.*

router. On Plane B the function is generic (the authors note the same effect in ZnO/DMSO colloids, ferrofluids and CNT liquid marbles), giving low $I_1 \approx 1–2$ bits and $I_2 \approx 10–15$ bits.

### S14. Worked estimate: Dúzs mechano-adaptive meta-gel [1]

A metamaterial strain-gate with an all-or-nothing threshold (sensor) feeds an autocatalytic urea–urease reaction-diffusion front at 0.11 mm·min$^{-1}$ (processor, with gain and long-range transmission), which drives global 7× stiffening via nanofibrillation or actuation via competitive swelling (actuator). It is the most architecturally complete of the four but open-loop (feedforward amplification, not regulatory feedback) and single-use (irreversible nanofiber write), so it stops short of closed-loop adaptation.

***Information and rate.*** A thresholded binary decision gives $I_3 \approx 1$ bit per event (the front distributes but does not add decision content). A complete sense → transmit → adapt event is set by front transit across a cm-scale device (~90 min at 0.11 mm·min$^{-1}$) plus actuation; taking $\nu \approx 3 \times 10^{-4}$ s$^{-1}$ (~1 event per hour) and $V \approx 10^{-7}$ m$^3$ (dogbone gauge ~35 µL, rheology cell ~120 µL) gives $\dot{I}_3 \approx 3 \times 10^{-4}$ bits·s$^{-1}$ and $\dot{I}_3/V \approx 3 \times 10^3$ bits·s$^{-1}$·m$^{-3}$ (range $10^2$–$10^4$).

***Power density (key steps).*** Driven by urea hydrolysis, [urea] ≈ 50 mol·m$^{-3}$ with $\Delta G \approx -30$ kJ·mol$^{-1}$; with most urea consumed over the ~1 h event, $P \approx (50$ mol·m$^{-3} \times 3 \times 10^4$ J·mol$^{-1})/(3.6 \times 10^3$ s$) \approx 4 \times 10^2$ W·m$^{-3}$ (range $10^2$–$10^3$, ±~1 decade).

***Landauer ratio and static plane.*** $P_{\min} = 2.85 \times 10^{-21} \times 3 \times 10^3 \approx 9 \times 10^{-18}$ W·m$^{-3}$, so $P/P_{\min} \approx 3 \times 10^{19}$ [25,27,28], placing it in the soft-matter band, despite the device being by far the most sophisticated of the four. The sophistication is carried by the gain (a sub-threshold local touch amplified to a 7× global stiffening) and by a high static selectivity $I_2 \approx 40–60$ bits, above the engineered-electronics band; $I_1 \approx 1–2$ bits. This illustrates the framework's central point that architectural complexity and Plane A position are largely decoupled: a one-shot, ~1-bit, hour-scale decision scores modestly no matter how sophisticated the kernel that produces it.